\documentclass{article}
\usepackage[parfill]{parskip}
\usepackage{epsfig}
\usepackage{amsfonts}
\usepackage{amsmath}
\usepackage{amssymb}
\def\X{X_d}
\def\Arg{{\rm Arg}}

\def\bC{{\bf C}}
\def\bR{{\bf R}}

\def\Im{\mathop{\rm Im}\nolimits}
\def\Re{\mathop{\rm Re}\nolimits}
\def\Rp{{\bf R}_+}

\def\ch{\mathop{\rm ch}\nolimits}
\def\sh{\mathop{\rm sh}\nolimits}

\def\SS{{\cal S}}
\def\TT{{\cal T}}

\def\wh{\widehat}

\def \veps{\varepsilon}

\def\half{{\scriptstyle{1 \over 2}}}
\def\interior#1{\setbox1=\hbox{$#1$}\rlap{$#1$}\kern0.4\wd1\raise1.1\ht1%
\hbox{$\scriptstyle \circ$}}
\def\bydef{\mathrel{\buildrel \hbox{\scriptsize \rm def} \over =}}
\def\boxit#1#2{\setbox1=\hbox{\kern#1{#2}\kern#1}%
\dimen1=\ht1 \advance \dimen1 by #1 \dimen2=\dp1 \advance \dimen2 by #1
\setbox1=\hbox{\vrule height\dimen1 depth\dimen2\box1\vrule}%
\setbox1=\vbox{\hrule\box1\hrule}%
\advance \dimen1 by .4pt \ht1=\dimen1 \advance \dimen2 by .4pt \dp1=\dimen2
\box1\relax}
\def\endprf{\raise .5ex\hbox{\boxit{2pt}{\ }}}

\def\ifundefined#1{\expandafter\ifx\csname#1\endcsname\relax}

\def\beq{\begin{equation}}
\def\endq{\end{equation}}
\def\beqa{\begin{eqnarray}}
\def\endqa{\end{eqnarray}}

\newcommand\n{\kappa}
\newcommand\w{{\cal W}}
\newcommand\ww{{W}}
\def\coupl{{\gamma}}
\newcommand\f{f}
\newcommand\ovl{\overline}
\newcommand\manifold{{\mathbb H}_{d}}
\newcommand\Lobad{{\mathbb H}_{d}}

\newcommand{\x}{{\rm n}}

\newcommand{\sd}{{\mathbb S}_{d-1}}

\newtheorem{theorem}{Theorem}[section]

\begin{document}

\title{Triangular invariants, three-point functions and particle stability on
the de Sitter universe}
\author{Jacques Bros$^1$, Henri Epstein$^2$, Michel Gaudin$^1$,
Ugo Moschella$^{3,4}$,
Vincent Pasquier$^1$ \\
\\$^1$Institut de Physique Th\'eorique,
CEA - Saclay, France\\
$^2$Institut des Hautes \'Etudes
Scientifiques, 91440 Bures-sur-Yvette \\
 $^3$Universit\`a dell'Insubria, 22100 Como,
Italia,
\\$^4$INFN, Sez. di Milano, Italia}
\maketitle
\begin{abstract}
We study a class of three-point functions on the de Sitter universe and on
the asymptotic cone.
A blending of geometrical ideas and analytic methods is used to compute some
remarkable integrals,
on the basis of a generalized
star-triangle identity living on the cone and on the complex de Sitter
manifold. We discuss an
application of the general results to the study of the stability of
scalar particles on the Sitter universe.
\end{abstract}

\section{Prologue}

The main result of this paper is the following formula:

\begin{equation}
 h_d(\n,\nu,\lambda) \bydef  \int_1^\infty
P^{-\frac{d-2}{2}}_{-\frac{1}{2} + i \n} (u)\,
{P^{-\frac{d-2}{2}}_{-\frac{1}{2} +
i\nu}(u)P^{-\frac{d-2}{2}}_{-\frac{1}{2} +
i\lambda}(u)}{(u^2-1)^{-\frac{d-2}4}} \ du =  \label{integral} \end{equation}
\begin{equation} =
\frac{2^{\frac d 2}}{(4\pi)^{\frac32}
\Gamma\left(\frac{d-1}{2}\right)} \frac{ \ \prod_{\epsilon,\epsilon',\epsilon''=\pm 1}
\Gamma\left(\frac{d-1}{4}
+\frac{i\epsilon \kappa+i\epsilon' \nu+i\epsilon''\lambda}{2}\right)}
{\left [\prod_{\epsilon=\pm 1}\Gamma\left(\frac{d-1}{2}+i\epsilon \n\right)
\right ]
\left [\prod_{\epsilon'=\pm 1}\Gamma\left(\frac{d-1}{2}+i\epsilon' \nu\right)
\right ]
\left [\prod_{\epsilon''=\pm 1}\Gamma\left(\frac{d-1}{2}
+i\epsilon'' \lambda\right)\right ]}
\label{Theformula}
\end{equation}

expressing the integral of a product of three  Legendre functions of the first 
kind \cite{bateman1} 
as a ratio of products of Euler Gamma functions.
This beautifully symmetric formula is not listed in any of the handbooks
on integrals of special functions
available to us and appears to be new.
The steps involved in the proof-computation also
give rise to many interesting quantities having  possibly geometrical 
interpretations
that we have not yet fully explored in their mathematical and  physical 
consequences.

The problem of computing the integral (\ref{integral}) originates
from the study \cite{Bros:2006gs,bem} of particle decays in a de Sitter universe.
In that context, the real parameters $\kappa$, $\nu$ and $\lambda$ are
related to the masses of the particles involved
in the decay process and $d$ is the (complex) dimension of the
de Sitter universe where the process takes place.
However, the methods and the results we are going to present  have presumably
a wider interest and range of applications.

The study of particle  decays in a de Sitter universe  was initiated by
O.~Nachtmann \cite{N} in 1968. He showed, in a very special case,
that while a Minkowskian particle can never decay into
heavier products, a de Sitter particle can, although this effect is
exponentially small in the de Sitter radius.

The subject has  acquired a greater physical interest with the advent of 
inflationary cosmology.
In particular, the idea that particle decays during the (quasi-)de Sitter phase
may have important consequences on the physics of the early universe 
has been suggested recently \cite{Boyanovsky:1996ab,Boyanovsky:2004gq}
(see also the related works \cite{Boyanovsky:2005px,Boyanovsky:2005sh}).
The mathematical and physical difficulties related to
the lack of time-translation symmetry of the de Sitter universe,
and more generally of non-static (cosmological) backgrounds, 
have been tackled in \cite{Boyanovsky:1996ab,Boyanovsky:2004gq} 
by using the Schwinger-Keldysh formalism, 
which is suitable for studying certain aspects of 
quantum physics of systems out of equilibrium.
However, the approach described 
in \cite{Boyanovsky:1996ab,Boyanovsky:2004gq}
necessitates the introduction of a practical notion of
lifetime of an unstable particle which is completely different 
from the definition commonly used in quantum physics. 

Actually the lack of a commutative symmetry group of spacetime translations 
renders the mathematics extremely complicated but does not prevent 
computing the inverse lifetime of an unstable de Sitter particle
according to the  usual definition, 
namely as the inclusive transition probability {\em per unit time} from
an initial state to every possible final state: this computation, 
initiated by Nachtmann long ago, has been performed
at first order in perturbation theory in \cite{Bros:2006gs,bem}. 
After writing the relevant perturbative amplitude, the computation of the
lifetime of the de Sitter unstable particle amounts to two essentially 
distinct and independent steps:
\begin{enumerate}
\item  taking the adiabatic limit of the infrared regularized inclusive 
amplitude,
i.e. removing the infrared cutoff coupling factor necessary  to make the
integral expressing the amplitutde converge;
    \item computing the so called ``phase space'' coefficient, a quantity
which only depends
on the masses of the particles involved in the decay process.
\end{enumerate}
The first step has been largely discussed in \cite{Bros:2006gs,bem} and the
resulting mathematical structures elucidated there. The second step reduces 
to computing the integral at thr RHS of (\ref{integral}).

The quantity $h_d(\n,\nu,\nu)$, relative to a decay into two identical
particles, has been computed in \cite{Bros:2006gs,bem}.
This special case already  exhibits some concrete mathematical difficulties,
and has been solved in a purely analytical way
by the use of Mellin transform techniques and the evaluation of a Barnes-type
integral \cite{Bros:2006gs,bem}.

The above method fails however to provide a solution for the general case of
the production of two non-identical particles, i.e.
fails to give a solution to the general integral (\ref{integral}). For odd
values of $d$ the Legendre functions of the first kind
reduce to trigonometric-type functions; in these cases a direct computation
of $h_d(\kappa,\nu,\lambda)$ is possible.
To give an example, one can solve the three-dimensional
problem ($d=3$) by  an elementary computation. Indeed for $d=3$ there holds
the particularly simple expression of
the Legendre function:
\begin{equation}P^{-\frac{1}{2}}_{-\frac{1}{2} +
i\nu}(\cosh v) =
\sqrt{\frac{2}{\pi \sinh v}}\,\frac{\sin \nu v}{\nu}.
\end{equation}
A straightforward computation then gives:
\begin{eqnarray} h_3(\kappa,\nu,\lambda)   &=&
\frac{1}{\sqrt{8  \pi} {\n\nu \lambda}}\ \frac{{\sinh (\pi \kappa )
\sinh (\pi \lambda ) \sinh (\pi  \nu )}}
{ \cosh\frac{ \pi (\kappa -\lambda-\nu)}{2}
\cosh\frac{ \pi (\kappa +\lambda-\nu)}{2} \cosh\frac{ \pi (\kappa -\lambda+\nu)}{2}
\cosh\frac{ \pi (\kappa +\lambda+\nu)}{2}}.
\end{eqnarray}
The general odd-dimensional case $d=2n+1$ can similarly be tackled by
(increasingly cumbersome) elementary integration techniques.
On the contrary, the computation of the even dimensional cases (including the
physically relevant four-dimensional de Sitter universe)  is very very
far from obvious.

Some of the geometric ideas necessary to overcome the difficulties of the
integral (\ref{integral}) were contained in an unpublished work
by one of us\footnote{The problem studied in \cite{gaudin} was the
construction of a generating function for the Clebsch-Gordan coefficients of
the group $SU(2)$ following a method originally due to Wigner.
Given three spinors $\psi_i = \left({\xi_i},{\eta_i}\right)\  i=1,2,3,$ one
can construct the unimodular invariants
$[\psi_i \psi_j] = \xi_i \eta_j - \xi_j \eta_i$; the quantity
$
I = [\psi_1 \psi_2]^{a_3}[\psi_2 \psi_3]^{a_1}[\psi_3 \psi_1]^{a_2}
$
is  an $SU(2)$-invariant polynomial trilinear in the vectors of the
representation
$D_{j_i}= {\xi_1^{j_i+m}\eta_1^{j_i-m}}/{\sqrt{(j_i+m)!(j_i-m)!}}=x_i(m)$,
where $a_1+a_2 = 2j_3$, etc..
The coefficients of the above polynomial are proportional to the Clebsch-Gordan
coefficients. To determine the constant $C$
it is sufficient to compute the following average on the unit sphere
\begin{equation}
J = \int |I|^2 d\Omega_1 d\Omega_2 d\Omega_3 =
C^2 \left(\begin{array}{ccc} j_1 & j_2 & j_3 \\ m_1 & m_2 & m_3
\end{array} \right)^2 = C^2
\end{equation}
with the parametrization $\xi =  \cos\frac \theta 2  \ e^{ i \frac \phi 2}$ and
$\eta =  \sin\frac \theta 2  \ e^{ -i \frac \phi 2}$.
Since $\left|[\psi_i \psi_j]\right|^2  = \sin^2 \frac {\gamma_{ij}} 2
= \left(\frac {\Delta n_{ij}} 2 \right)^2 $ one can recognize here the
triangular invariant introduced in Section (\ref{st1}) and computed in
Section (\ref{st2}). Here the calculation of the invariant $J$ is done for
non-integer values of the exponents (non-compact case). The method we will use
generalize the partial integrations that were enough to solve the integer case
in \cite{gaudin} by using the fractional calculus.
} \cite{gaudin}.
Combining those ideas with a the geometrical properties of the complex de
Sitter manifold
provides a way to solve the problem. The result is displayed in Eq. (\ref{Theformula}).

Beyond the  study of the  of de Sitter particle decays,
there are other potential applications of the formula
(\ref{Theformula}) and of the methods used to derive it which
include  the study of tensor product of representations of non-compact groups,
many new integral relations involving products of hypergeometric
functions, other applications to de Sitter and/or anti de Sitter QFT etc..

\section{Legendre functions and de Sitter Klein-Gordon fields: a short review}
\label{legendresec}
The computation of the integral (\ref{integral}) requires several steps in
which the geometrical features  of the complex de Sitter manifold  enter
in a crucial way.
The first important step consists in returning to the meaning of the
Legendre functions of the first kind $P(u)$ as
{\em two-point functions} of quantum fields on
a complexified de Sitter spacetime
\cite{Bros:1990cu,Bros:1994dn,Bros:1995js,Bros:1998ik}.
The variable  $u$, appearing as integration variable in the r.h.s. of
(\ref{Theformula}),
is understood as a geometrical invariant  $u = u(x,x')$  relating two points
$x,x'$ of a (complex) de Sitter hyperboloid. This idea allows in particular
a natural way for understanding many of the mathematical properties of the
Legendre functions and gives also a simple procedure to build many of their
integral representations.
Here follows a short account of the construction.

Consider a $(d+1)$-dimensional Minkowski spacetime
${M}_{d+1}$; an event $x$ is parameterized by a set of
inertial coordinates $x^0,\ldots x^{d}$; the scalar product of
two events of ${ M}_{d+1}$ is the Lorentz-invariant product
$
x\cdot {x'} = x^{0}{x'^{0}}-{x^{1}}{x'^{1}}-\ldots-
{x^{d}}x'^{d}. $ The $d$-dimensional de Sitter spacetime is
represented as the one-sheeted hyperboloid
\begin{equation}
\X  = \{x\in {M}_{d+1} : x\cdot x = x^2=  -R^2\}
\end{equation}
embedded in ${ M}_{d+1}$. The Lorentzian geometry of the
de Sitter manifold is induced by the causal structure of the
ambient spacetime:
\begin{eqnarray}
V^+ &=& \{\xi \in { M}_{d+1}:\xi^2 =  \xi \cdot \xi >0 ,\
\xi^0>0\},\\
C^+ &=& \{\xi \in {M}_{d+1}:\xi^2 =  \xi \cdot \xi =0 ,\
\xi^0>0\};
\end{eqnarray}
the future cone $V^+$ of the ambient spacetime induces the Lorentzian
global causal ordering on the de Sitter
universe: $x$ is in the future of $x'$ if and only if $x-x' \in V^+$.
The forward light-cone
$C^+$ of the ambient spacetime also plays the role of
the space of momentum directions in de Sitter momentum space
 \cite{Bros:1995js, Cacciatori:2007in}. The de Sitter
invariance group is the Lorentz group of the ambient spacetime $SO(1,d)$.

A de Sitter generalized free field  $\phi$
is fully characterized by its two-point vacuum expectation
value $\w(x, x')$ which is assumed to be  be a local and de Sitter
invariant distribution.
Since there is no global de Sitter energy operator, a true spectral
condition does not exist in the de Sitter spacetime;
there is however a suitable replacement  that
can be formulated \cite{Bros:1995js} by moving to the complex de Sitter
manifold
\begin{equation} \X^{(c)}  = \{z\in { M}_{d+1}^{(c)} : z\cdot z =
-R^2\}
\end{equation}
and requiring that $\w(x, x')$ be the boundary value on the
reals of a de Sitter invariant function $\ww_m(z,z')$ holomorphic in the tubular
domain ${\cal T}^-\times {\cal T}^+$ with slow increase properties at infinity,
where
\begin{equation}
{\mathcal T}^\pm = \{{(M}_{d+1}\pm iV^+)\cap \X^{(c)}\}=
\{ z=x+iy \in \X^{(c)},\  y\cdot y >0, \ {\rm sign} (y^0) =\pm\}.
\end{equation}
de Sitter invariance can then be used to show that $\ww(z,z')$
is actually maximally analytic, i.e. it is  analytic in the
domain
\begin{equation}
\Delta = \{(z,z') \in\X^{(c)}\times \X^{(c)}: \  (z-z')^2 \not \in {\bf R}^+\}.
\end{equation}
For a thermodynamical interpretation of the above analyticity property,
see \cite{Bros:1995js,Bros:1998ik,Gibbons:1977mu}.
By introducing the de Sitter invariant variable
\begin{equation}
\zeta = \frac{z \cdot z'}{R^2} \
(\ = u \makebox{ when real and greater than one})
\end{equation}
there holds a simple description of $\ww$ in terms of a
function $w$ of the single variable $\zeta$, namely
$w(\zeta) =  W(z,z')$, holomorphic in
image of the domain $\Delta$
\begin{equation}
\Pi =  \{ \zeta \in \bC , \ \ \zeta \not = -1 - {\bf R^+}\}.
\end{equation}
For a Klein-Gordon field $\phi$ with mass $m \ge
0$ the two-point function must also be a bisolution of the Klein-Gordon equation
\begin{equation} (\Box_x +m^2)\w_m(x, x')= (\Box_{x'}
+m^2)\w_m(x, x') = 0
\end{equation}
where $\Box$ is the Laplace-Beltrami operator relative to the
de Sitter geometry.
It is useful to introduce
a dimensionless parameter $\nu$ related to the mass $m$ as follows:
\begin{eqnarray}
m^2R^2 &=& \left ( {d-1 \over 2} \right )^2 + \nu^2
\label{w.2}.
\end{eqnarray}
By abuse of language we will call $\nu$ a mass parameter even if it is dimensionless. Given a complex $\nu$ the corresponding two-point function $\ww_{m}(z,z')= \ww_\nu(z,z') = w_\nu(\zeta)$ is written in terms of
Legendre functions of the first kind as follows:
\begin{eqnarray}
 \ww_\nu(z,z') = \frac{ \Gamma\left(\frac{d-1}{2} +i
\nu\right)\Gamma\left( \frac{d-1}{2} -i
\nu\right)}{2(2\pi)^{\frac{d}{2}} R^{d-2}}
(\zeta^2-1)^{-\frac{d-2}{4}}\,
P^{-\frac{d-2}{2}}_{-\frac{1}{2} + i\nu}\left(\zeta\right)
\label{legendre}
\end{eqnarray}
$z,z'$ are events belonging to $\Delta$; the normalization ensures that the canonical
commutation relations hold with the correct coefficient.

The range $m \geq  m_c = (d-1)/2R $ corresponds to the {\em
principal series} of unitary irreducible representations of the
de Sitter group ($\nu$ real) while $0 < m < m_c$ corresponds
to the { complementary series} ($\nu$ imaginary). These
restrictions ensure that the boundary value $\w_m$ is positive definite and
therefore a quantum theoretical interpretation is available.
Note also the symmetry property
\begin{equation}
\w_\nu(x,x') =
\w_{-\nu}(x,x')
\ \ \ \left({\rm implied}\ \ {\rm by}\ \ \
P^{-\frac{d-2}{2}}_{-\frac{1}{2} + i\nu}(\zeta)=
P^{-\frac{d-2}{2}}_{-\frac{1}{2} - i\nu}(\zeta) \right),
\label{sym}
\end{equation}
which holds for all  $\nu$
and will play a role in one structural aspect of the derivation of formula (\ref{integral}).

\subsection{Plane waves expansion of Legendre functions}

There exists a Fourier-type representation of the two-point
functions (\ref{legendre}) which is of fundamental importance to understand the
above properties, and generally speaking to understand de
Sitter QFT. It is constructed by using a natural  basis of
plane-wave solutions $\psi_\nu$ of the Klein-Gordon equation
\begin{equation} (KG)_\nu\ \psi_\nu(z) = \left[\Box_z   +
\left( {d-1 \over 2R} \right)^2  + \left(\nu\over R\right)^2 \right ]
\psi_\nu (z) =0,
\label{KGnu}
\end{equation}
which are
parameterized by the choice of a lightlike vector $\xi\in C^+$ as follows:
\begin{equation}
\psi_\nu(z,\xi) = \left(\frac{ z\cdot \xi}{R}\right)^{-\frac{d-1}2 + i \nu}. \label{waves}
\end{equation}
These waves are well-defined and analytic in each of the tubes
$\cal T^+$ and $\cal T^-$. Then, for $z \in \TT^-$ and $z'
\in \TT^+$ the following Fourier-type (i.e. momentum space)
representation of the two-point function holds true:
 \begin{equation} \w_\nu(z,z') =
{\Gamma({d-1\over
2}+i\nu)\Gamma({d-1\over 2}-i\nu) e^{-\pi\nu} \over 2^{ d+1}
\pi^d R^{d-2}}\,\int_{\gamma}\left({z\cdot \xi}\right)^{-{d-1\over 2}
+ i\nu} \left({\xi\cdot z'}\right)^{-{d-1\over 2} -
i\nu}\alpha(\xi) , \label{f14}
\end{equation}
In  standard coordinates
the ($d-1$)-form $\alpha(\xi)$ is written
\begin{equation} \alpha(\xi) = (\xi^0)^{-1} \sum_{j=1}^d (-1)^{j+1}
\xi^j\,d\xi^1\ldots\ \wh{d\xi^j}\ldots\ d\xi^d\ .
\label{f.16}\end{equation}
In (\ref{f14}), the $(d-1)-$form under the integration sign is closed and $\gamma$
denotes any
($d-1$)-cycle in the forward
light-cone $C^+$ which is homologous to the following cycle  $\gamma_0$: the support
of $\gamma_0$ is represented as follows by the unit sphere,
$\sd$ (equipped with its canonical orientation):
\begin{equation}
{\rm supp}\gamma_0=\sd= C^+ \cap \{\xi\ :\ \xi^0 =1\}
=  \{ \xi \in C^+ : {\xi^1}^2 + \ldots + {\xi^d}^2 =
1\}.
\end{equation}
With this choice $\alpha(\xi)$  coincides
with the rotation invariant measure on  $\sd$ normalized as
follows:
\begin{equation}
\omega_{d}=\int_{\gamma_0}\alpha(\xi) = \frac{2\pi^\frac d2}{\Gamma\left(\frac d2\right)}.
\label{norms}
\end{equation}
\subsection{Lobatchevski space and a remarkable representation of Legendre functions} \label{loba}
A specially important parametrization of the Fourier-type
representation  is obtained by evaluating  (\ref{f14}) at the purely imaginary
events $z=0-iy\in {\cal T}^-$ and $z = 0+ iy'\in {\cal T}^+$; $y$
and $y'$ can be visualized as points belonging to a  Lobatchevski
space, modeled as the upper sheet of a two-sheeted hyperboloid:
\begin{equation}
\manifold  = \{y\in{\mathbb M}_{1,d}:\ y^2 = y\cdot y = R^2,\;\; y^{0}>
0\}.
\end{equation}
We will make use of  the following  spherical parametrization of
$\Lobad$:
\begin{equation}
y(u,\x)= R(u, \x^1 \sqrt{u^2 - 1},\ldots, \x^d\sqrt{u^2 - 1}) 
\end{equation}
where $u\geq 1$ and ${ \x} \in \sd$; in these coordinates the Lorentz-invariant
measure $dy$ is written
\begin{equation}
dy = R^d(u^2-1)^{\frac{d-2}2}du d\x \label{measureinv}
\end{equation}
where $d\x$ denotes the rotation-invariant measure on the
sphere ${\mathbb{S}}_{d-1}$ normalized as in Eq. (\ref{norms}).
With the above
specifications, Eq. (\ref{f.14}) allow us to write:
\begin{eqnarray} \w_\nu(-iy,iy') =
{\Gamma({d-1\over 2}+i\nu)\Gamma({d-1\over 2}-i\nu)  \over 2^{
d+1} \pi^d} \int_{\gamma}\left({y\cdot \xi}\right)^{-{d-1\over
2} + i\nu} \left({\xi\cdot y'}\right)^{-{d-1\over 2} -
i\nu}d\mu_\gamma(\xi) = \label{f.14}\cr = \frac{
\Gamma\left(\frac{d-1}{2} +i \nu\right)\Gamma\left(
\frac{d-1}{2} -i \nu\right)}{2(2\pi)^{\frac{d}{2}}}
\left(\left({y\cdot
y'}\right)^2-1\right)^{-\frac{d-2}{4}}\,
P^{-\frac{d-2}{2}}_{-\frac{1}{2} + i\nu}\left( {y\cdot
y'}\right).
\end{eqnarray}
Here and in the following we have set $R=1$;
by choosing in particular $\gamma =\gamma_0$ and $y' =
(1,0,\ldots,0)$ so that ${y\cdot y'}= y^0 = u\geq 1$, we then get the
following integral representation:
\begin{eqnarray}
\left(u^2-1\right)^{-\frac{d-2}{4}}\, P^{-\frac{d-2}{2}
}_{-\frac{1}{2} + i\nu}\left(u\right)\ = \frac{1 } {
(2\pi)^\frac d2 } \int_{\gamma_0} \left({y \cdot \xi
}\right)^{-{d-1\over 2} - i\nu}\alpha(\xi).
\label{formulaP}\end{eqnarray} This formula will be of crucial importance
for computing
$ h_d(\n,\nu,\lambda)$, since it allows one to rewrite
the integral in Eq. (\ref{Theformula}) as
the following multiple integral over the manifold $\manifold  \times \sd \times
\sd \times \sd$:
\begin{eqnarray}
&& h_d(\n,\nu,\lambda)= \cr &&
\frac{1 } {(2\pi)^\frac {3d}2 \omega_{d-1}}
\int_{\gamma_0} \int_{\gamma_0} \int_{\gamma_0}
\int_{\manifold} \left({y \cdot
\xi_1}\right)^{-{d-1\over 2} - i\n} \left({y \cdot
\xi_2}\right)^{-{d-1\over 2} - i\nu} \left({y \cdot
\xi_3}\right)^{-{d-1\over 2} - i\lambda}dy \,
\alpha(\xi_1)\,\alpha(\xi_2)\,\alpha(\xi_3)\label{res}  \cr  &&\end{eqnarray}
where we have used the measure (\ref{measureinv}) and the normalization (\ref{norms}).

\subsection{K\"all\'en-Lehmann-type representation for general
two-point functions}

Consider again a general two-point function such that its reduced form $w(\zeta)$
is analytic in the cut-plane $\Pi$ and
uniformly bounded at infinity by a certain power
$|\zeta|^{m_0}$. It has been shown in \cite{bv} that for  $-1<m_0 <0$
there exists  an integral representation of  $w(\zeta)$ the following form:
\begin{eqnarray}
w(\zeta) & = &
\frac 1 {2 (2\pi)^{\frac{d}{2}}}\int_{-\infty}^{\infty}
\ \left(m_0 + {d-1\over 2} +i\kappa\right)\
{ \Gamma\left(m_0 + d-1 +i
\kappa\right)\Gamma\left( -m_0 -i
\kappa\right)} \ \times
 \cr
&&  \times \ G(m_0+i\kappa)\
(\zeta^2-1)^{-\frac{d-2}{4}}\,
P^{-\frac{d-2}{2}}_{m_0+ {d-2\over 2} + i\kappa}
(\zeta)
d\kappa ;
\label{preKL}
\end{eqnarray}
the function $G(m_0 +i\kappa)$
is the boundary value of a  function
$G(s)$  holomorphic in the half-plane $\Re s> m_0$.
$G$ is obtained as a Laplace-type transform of the discontinuity
$\Delta w(\zeta)$ of $w(\zeta)$ across the cut $]-\infty,-1]$ (we will not need the explicit expression given in \cite{bv}, Eqs.
III 10 and III 11).

The  results of \cite{bv} can be extended to
the case $m_0= -{d-1\over 2}$, which is  relevant for de Sitter quantum field
theory, because, in that case,
the Legendre functions involved in (\ref{preKL}) are
all the free-field two-point
functions  of the principal series.
We omit the details of the proof of formula (\ref{preKL}) under the assumption that
$|w(\zeta)|$ is bounded by
$|\zeta|^{-{d-1\over 2}}$.
Inserting the value
$m_0=-{d-1\over 2}$
in Eq. (\ref{preKL})
and taking the symmetry condition (\ref{sym}) into
account and puts
, one obtains:
\beq
w(\zeta) =
\int_{0}^{\infty} \rho(\kappa)
{ \left|\Gamma\left(\frac{d-1}{2} +i
\kappa\right)\right|^2
\over 2(2\pi)^{\frac{d}{2}}}
(\zeta^2-1)^{-\frac{d-2}{4}}\,
P^{-\frac{d-2}{2}}_{-\frac{1}{2} + i\kappa}
(\zeta)
d\kappa^2  = \int_0^{\infty}
\rho(\kappa)\ w_\kappa(\zeta) d\kappa^2
\label{KL} .
\endq
which is a genuine K\"all\'en-Lehmann-type representation of $w(\zeta)$ with weight
\begin{equation}
\rho(\kappa)=  \frac{G\left(-{d-1\over 2} -i\kappa\right)
-G\left(-{d-1\over 2} +i\kappa\right)  }{2i} \label{laplinv}
\end{equation}
The computation the K\"all\'en-Lehman weight $\rho$
can also be tackled by invoking the generalized Mehler-Fock transformation theory \cite{MagnusOS}, as we do here.
Eq. (\ref{laplinv}) takes the following concrete form:
\beq
\rho(\kappa) = 2\  (2\pi)^{d-2\over 2}
\ \sinh \pi \kappa
\int_1^{\infty}w(\zeta)\
P^{-\frac{d-2}{2}}_{-\frac{1}{2} +
i\kappa}(\zeta)\ (\zeta^2-1)^{d-2\over 4} \ d\zeta.
\label{MF}
\endq

\section{Decay of de Sitter unstable particles}

The study of particle disintegration in the de Sitter universe
has been initiated in a pioneering paper by Nachtmann.

Consider three independent neutral Klein-Gordon
scalar fields $\phi_0,\ \phi_1, \ \phi_2$ with real  mass parameters  $\kappa,
\nu,\lambda$ respectively (i.e. the fields  in the principal series) and
an interaction term of the form
\[
\int \coupl\,g(x)\,{\cal L}(x)\,dx,\ \ \ {\cal L}(x) =\
:\phi_0(x)\phi_1(x)\phi_2(x): \label{1.3.3}\] where $g$ is a
smooth spacetime dependent "switching-on factor" which, in the
end, should be made to tend to the constant 1.
Self-interactions ${\cal L}(x) =\ :\phi(x)^{3}:$ are a special
case of this coupling. Let us consider the decay process
\begin{equation}
0\to 1+2
\end{equation}
Let in particular $\Psi_0$ be a one-particle state of the form
\[ \Psi_0 =  \int \f(x)\,\phi_0(x)\Omega\,dx ;\]
the smooth test function $\f(x)$ contains the physical details
about the preparation of the quantum state of the unstable
particle whose disintegration we aim to study. The following
general formula for the transition probability holds true
\cite{Bros:2006gs}:
\begin{eqnarray} && {\Gamma(0;1,2)=
{\coupl^2\,2\pi |\coth \pi \kappa| \,\int g(x)\,|F(x)|^2\,dx \over \int
\ovl{\f(x)}\w_{\kappa}(x,\ y)\,\f(y)\,dx\,dy}}\,\rho_{\nu,\lambda}(\kappa).\;\;\;\;
\label{1.17}\end{eqnarray}

Here the convolution $F(x) = \int \w_{\kappa}(x,\ y)\,\f(y)\,dy$ is the
"positive-frequency" solution of the KG equation with mass
$\kappa$ associated with the test-function $\f$; the denominator
is the squared norm of $\Psi_0$. Note that the first factor in
this formula does not depend on the  the decay particles but
only on the wavefunction of the incoming unstable particle. The
infrared problem is contained in this factor and has to be
overcome when letting the remaining $g(x)$ tend to 1 (adiabatic
limit). We will not treat this problem here and refer to
\cite{bem}.

The second factor is the relevant K\"all\'en-Lehmann weight of
the bubble diagram corresponding to two-point function of a composite field,
obtained
as the Wick product of the Klein-Gordon fields with mass parameters
$\nu$ and $\lambda$:
\beq
w(\zeta) =w_\nu(\zeta)w_\lambda(\zeta). \label{uuu}
\endq
This two-point function
is well-defined and analytic in the cut-plane $\Pi$.
Moreover, for real values of $\lambda$ and $\nu$  (i.e. for fields belonging
to the principal series)
it is bounded in $\Pi$ by $|\zeta|^{-{d-1}}$
and therefore, a fortiori,  by
$|\zeta|^{-{d-1\over 2}}$.  In particular, the Laplace-type  transform
$G_{\nu,\lambda}(s)$ of $(\ref{uuu})$
is analytic in the half-plane $\{s\in \bC;\ \Re s> -(d-1)\}$.
Thus, there exists a K\"all\'en-Lehmann representation
(\ref{KL}) of $w_\nu(\zeta)w_\lambda(\zeta)$:
\beq
w_\nu(\zeta) w_\lambda(\zeta)  = \int_0^{\infty}
\rho_{\nu,\lambda}(\kappa)w_\kappa(\zeta)
d\kappa^2 = \int_\bR \kappa \rho_{\nu,\lambda}(\kappa)w_\kappa(\zeta)\,d\kappa.
\label{KLwick}
\endq
The weight $\rho$, as given in Eq. (\ref{laplinv}) inherits analyticity
properties from the aforementioned properties of
$G_{\nu,\lambda}(s)$ of $(\ref{uuu})$ and
is itself holomorphic in the strip
$\{s\in \bC:\  -(d-1)< \Re s <0\}$ and therefore {\sl it cannot
vanish on any open interval of the line} $s= -{d-1\over 2}+i\kappa$.

This immediately implies that, in de Sitter spacetime, there is nothing such as
the "subadditivity condition" of
the Minkowski case: that property would require that
$\rho_{\nu,\lambda}(\kappa)$ should vanish if
$\kappa< \nu + \lambda$ and this would forbid
the decay of a particle of mass $\kappa$
into a pair of particles of masses
$\nu$ and $\lambda$ when
$\kappa< \nu + \lambda$.
In contrast, in the de Sitter universe
the disintegration of a particle of a given mass can give
rise to two heavier particles if such a coupling enters in the
interaction Lagrangian.

In the following sections we will explicitly compute the K\"all\'en-Lehmann
weight  by a mixture of geometrical insights and analytical techniques.
By inserting the \ref{legendre} of
in the Mehler-Fock transform  (\ref{MF}) of $w_\nu w_\lambda$
it follows that
$\rho_{\nu,\lambda}(\kappa)$ is
proportional to the integral (\ref{Theformula}); more precisely, one obtains:
\begin{eqnarray}
\rho_{\nu,\lambda}(\kappa) &=& \frac{
\Gamma\left(\frac{d-1}{2}+i \nu\right)
\Gamma\left(\frac{d-1}{2}-i \nu\right)
\Gamma\left(\frac{d-1}{2} +i\lambda\right)
\Gamma\left(\frac{d-1}{2} -i\lambda\right)
}{2(2\pi)^{1+{d\over 2}}}
\sinh (\pi \kappa) \  h_d(\n,\nu,\lambda),
\label{mf}\end{eqnarray}

\section{A special class of de Sitter three-point functions.}

The general properties of de Sitter two-point functions can, to some extent,
be generalized to
a class ${\cal C}_{F}$ of three-point functions $W(x_1,x_2,x_3)$  on de Sitter
spacetime, such that
\begin{enumerate}
\item  $W(x_1,x_2,x_3)$ is a distribution on
$X_d\times X_d\times X_d$ which is decomposable as a sum of two
boundary values of holomorphic functions $W_\varepsilon(z_1,z_2,z_3),$
from the respective tubular
domains ${\cal T}^-\times {\cal T}^{\varepsilon}\times
{\cal T}^+$
where $\varepsilon=
+\ {\rm or} \ -. $

\item Each  function
$W_\varepsilon(z_1,z_2,z_3)$ is invariant under the complex de Sitter
group $SO_0(1,d)^{(c)}$ and therefore it
coincides with a holomorphic function of the three complex invariants
$z_i\cdot z_j = \zeta_{ij}, \ i,j = 1,2,3,$ (since $z_i^2= -1$) in the
image of the corresponding tubular
domain.

\item  Each distribution
$W_\varepsilon(x_1,x_2,x_3)$ admits a
Fourier-type transform on the one-sheeted hyperboloid defined in terms of
the plane waves (\ref{waves})
(see also \cite{Bros:2003uw} where the case $d=2$ has been treated in detail).
\end{enumerate}
We do not expect that the above properties hold for general interacting
quantum field theories; in particular they do not apply
to the general class introduced in \cite{Bros:1998ik}.
They can be however useful in a perturbative context.

The transform of
$W_\varepsilon(x_1,x_2,x_3)$
is  a distribution on $(C^+)^3\times \bR^3$ defined by
\begin{eqnarray}
\tilde W_\varepsilon(\xi_1,\xi_2,\xi_3; \kappa,\nu,\lambda)=
\int_{X_d^3}
\left({x_1 \cdot
\xi_1}\right)_+^{-{d-1\over 2} - i\kappa} \left({x_2 \cdot
\xi_2}\right)_{-\varepsilon}^{-{d-1\over 2} - i\nu} \left({x_3 \cdot
\xi_3}\right)_-^{-{d-1\over 2} - i\lambda} \times \cr \times
W_\varepsilon(x_1,x_2,x_3)
\ dx_1\ dx_2\ dx_3.
\label{FH}
\end{eqnarray}
Conversely, each holomorphic function
$ W_\varepsilon$ is recovered in its
respective domain ${\cal T}^-\times {\cal T}^{\varepsilon}\times
{\cal T}^+$
by an inversion formula,
which includes an appropriate weight-function $\sigma_d$ on $\bR^+$:
\begin{eqnarray}
&& W_\varepsilon(z_1,z_2,z_3)=
\int_{{\bf R}_+^3}
\sigma_d(\kappa)
\sigma_d(\nu)
\sigma_d(\lambda)
\ d\kappa d\nu d\lambda \
\times \cr && \times \
\int_{\gamma_0^3}
\left({z_1 \cdot
\xi_1}\right)^{-{d-1\over 2} + i\kappa} \left({z_2 \cdot
\xi_2}\right)^{-{d-1\over 2} + i\nu} \left({z_3 \cdot
\xi_3}\right)^{-{d-1\over 2} + i\lambda}
\tilde W_\varepsilon(\xi_1,\xi_2,\xi_3; \kappa,\nu,\lambda)
\alpha(\xi_1)
\alpha(\xi_2)
\alpha(\xi_3).
\label{FHI}  \cr &&
\end{eqnarray}
$\tilde W_\varepsilon$ depends on $(\xi_1,\xi_2,\xi_3)$ only through the (Lorentz) invariants
$\xi_1\cdot\xi_2,$ $
\xi_2\cdot\xi_3,$ $
\xi_3\cdot\xi_1,$ (since $\xi_j^2 =0$). The homogeneity properties of  (\ref{FH})
w.r.t the variables $\xi_j \in C^+$ imply that
\beq
\sigma_d(\kappa)
\sigma_d(\nu)
\sigma_d(\lambda)\
\tilde W_\varepsilon(\xi_1,\xi_2,\xi_3; \kappa,\nu,\lambda)=
\hat\rho_\varepsilon(\kappa,\nu,\lambda)\
(\xi_1\cdot\xi_2)^{a_3}\
(\xi_2\cdot\xi_3)^{a_1}\
(\xi_3\cdot\xi_1)^{a_2}.
\label{tildew}
\endq
where we have introduced the parameters
\beq
a_1= -{d-1\over 4} -i{\nu +\lambda -\kappa\over 2},
\ a_2= -{d-1\over 4} -i{\lambda +\kappa-\nu \over 2},
\ a_3= -{d-1\over 4} -i{\kappa +\nu -\lambda \over 2}, \label{param}
\endq
The inversion formula (\ref{FHI}) can therefore be rewritten as follows:
\beq
W_\varepsilon(z_1,z_2,z_3)=
\int_{{\Rp}^3}
\hat \rho_\varepsilon(\kappa,\nu,\lambda)\
w_{\kappa,\nu,\lambda}(z_1,z_2,z_3)
\ d\kappa d\nu d\lambda  \label{KL3}
\endq
where
\begin{eqnarray}
 w_{\kappa,\nu,\lambda}(z_1,z_2,z_3) &=&
\int_{\gamma_0^3}
\left({z_1 \cdot
\xi_1}\right)^{-{d-1\over 2} + i\kappa} \left({z_2 \cdot
\xi_2}\right)^{-{d-1\over 2} + i\nu} \left({z_3 \cdot
\xi_3}\right)^{-{d-1\over 2} + i\lambda} \times \cr
&&  \times \
(\xi_1\cdot\xi_2)^{a_3}\
(\xi_2\cdot\xi_3)^{a_1}\
(\xi_3\cdot\xi_1)^{a_2}
\alpha(\xi_1)
\alpha(\xi_2)
\alpha(\xi_3).
\label{FHI+}
\end{eqnarray}
The three-point
function $w_{\kappa,\nu,\lambda}(z_1,z_2,z_3)$ manifestly
satisfies the triplet of Klein-Gordon equations:
\beq
[(KG)_\kappa]_{z_1}
w_{\kappa,\nu,\lambda}(z_1,z_2,z_3)
= [(KG)_\nu]_{z_2}
w_{\kappa,\nu,\lambda}(z_1,z_2,z_3)
= [(KG)_\lambda]_{z_3}
w_{\kappa,\nu,\lambda}(z_1,z_2,z_3)
=0  \label{KG3}
\endq
in the (non-connected) complex open set $(z_1,z_2,z_3)\in
{\cal T}^{\pm}\times {\cal T}^{\pm}\times
{\cal T}^{\pm}$ where it is holomorphically defined via Eq. (\ref{FHI+}).
This set contains in
particular the relevant tubular domains
${\cal T}^-\times {\cal T}^{\varepsilon}\times
{\cal T}^+$
in which the integral representation (\ref{KL3})
is meaningful.
Formula (\ref{KL3}) has the shape of
generalized K\"all\'en-Lehmann representation for all the three-point functions
which belong to the class ${\cal C}_{F}$ on the basis of three-point
functions  satisfying the Klein-Gordon  system (\ref{KG3}),

\section
{A star-triangle relation and a class of triangular invariants on the hypersphere}
\label{st1}
Before the computation of $h_d(\kappa,\nu,\lambda)$ can be made possible
we need to introduce two further ingredients: a generalized star-triangle relation and a class of triangular invariants on the hypersphere.
They both come out from the study of
 the following integral on the Lobachevski manifold  $y\in \manifold$ (see Sec. \ref{loba}):
\begin{eqnarray}
F_{a_1,a_2,a_3}(\xi_1,\xi_2,\xi_3) &=&
\int_{\manifold} (y\cdot \xi_1)^{a_2+a_3}(y\cdot \xi_2)^{a_3+a_1}(y\cdot \xi_3)^{a_1+a_2} dy =
\label{defF} \\ &=&
c(a_1,a_2,a_3) (\xi_1\cdot \xi_2)^{a_3}
(\xi_2\cdot \xi_3)^{a_1}(\xi_3\cdot \xi_1)^{a_2} \label{st}
\end{eqnarray}
the second equality follows again from  Lorentz (i.e. de Sitter) invariance
and from the homogeneity properties of
(\ref{defF}) with respect to the variables $\xi_j$'s; the constant $c(a_1,a_2,a_3)$ remains to be determined.
This identity is a sort of  ``star-triangle relation'' with one important difference w.r.t. what is usually called ``star-triangle'':  the center of the star is a point of  $\manifold$ while the legs belong to the asymptotic cone $C^+$ i.e. the center and the legs  of the star do not belong to the same manifold.
By integrating both sides over the spherical basis ${\gamma_0}$ of the cone we get
\begin{eqnarray}
f(a_1,a_2,a_3) &=&
\int_{\gamma_0} \int_{\gamma_0} \int_{\gamma_0}
F_{a_1,a_2,a_3}(\xi_1,\xi_2,\xi_3)
\alpha(\xi_1)\,\alpha(\xi_2)\,\alpha(\xi_3) \nonumber
\\
&=& c(a_1,a_2,a_3)
\int_{\gamma_0}\int_{\gamma_0}\int_{\gamma_0}
(\xi_1\cdot \xi_2)^{a_3}
(\xi_2\cdot \xi_3)^{a_1}(\xi_3\cdot \xi_1)^{a_2}\
\alpha(\xi_1)\,\alpha(\xi_2)\,\alpha(\xi_3)
\label{fF} \\
&=& c(a_1,a_2,a_3) \times \hat J(a_1,a_2,a_3)
\label{fcJ}
\end{eqnarray}
The integral at the r.h.s.  is proportional to the value of the  three-point function
$w_{\kappa,\nu,\lambda}(z_1,z_2,z_3)$ at the special complex event
$z_1=z_2=z_3 = iy_0\in {\cal T}^+, \ y_0= (1,0\ldots,0)$:
\beq
w_{\kappa,\nu,\lambda}(iy_0,iy_0,iy_0) = e^{\pi(\kappa+\nu +\lambda)\over 2}
e^{-3(d-1)\pi i\over 4}
\int_{\gamma_0^3}
(\xi_1\cdot\xi_2)^{a_3}\
(\xi_2\cdot\xi_3)^{a_1}\
(\xi_3\cdot\xi_1)^{a_2}
\alpha(\xi_1)
\alpha(\xi_2)
\alpha(\xi_3).
\label{FHI0}
\endq
{}From Eqs (\ref{res}), (\ref{param}) and  (\ref{FHI0}), it follows that
computing $ h_d(\kappa,\nu,\lambda)$ is equivalent to integrating the
 star-triangle relation (\ref{fF})
w.r.t. the external legs:
\begin{eqnarray}
h_d(\kappa,\nu,\lambda)&=&
{1\over (2\pi)^{3d\over 2}\ \omega_{d-1}}\
f(a_1,a_2,a_3)= \label{hdw1}\cr &=&
{1\over
(2\pi)^{3d\over 2}\ \omega_{d-1}}\ e^{-\pi(\kappa+\nu +\lambda)\over 2}
e^{3(d-1)\pi i\over 4}\
w_{\kappa,\nu,\lambda}(iy_0,iy_0,iy_0)
\times c(a_1,a_2,a_3).
\label{hdw}
\end{eqnarray}

Define
\beq
\hat J(a_1,a_2,a_3)
=\int_{{\gamma_0}^3}
(\xi_1\cdot \xi_2)^{a_3}
(\xi_2\cdot \xi_3)^{a_1}(\xi_3\cdot \xi_1)^{a_2}\
\alpha(\xi_1)\,\alpha(\xi_2)\,\alpha(\xi_3),
\label{hatJ}
\endq
so that Eq. (\ref{fF}) is rewritten as follows:
\beq
f(a_1,a_2,a_3)= c(a_1,a_2,a_3) \ \hat J(a_1,a_2,a_3).
\label{fcJ1}
\endq
The integral (\ref{hatJ})
has a beautiful geometrical interpretation as a
triangular invariant on the hypersphere  \cite{gaudin}
$ \sd$. Consider indeed  the squared distance  $\Delta \x_{ik}^2$
between two points $\x_i$ and $\x_k$ belonging to $\sd$. The
Lorentzian scalar product of two points $\xi_i = (1,\x_i)$ and
$\xi_k=(1,\x_k)$ belonging to the spherical cycle $\gamma_0$ of
the forward lightcone $ C^+$ is proportional to the squared
distance $\Delta \x_{ik}^2$:
\begin{equation}
 { \Delta \x_{ik}^2} =   {(\x_i-\x_k)^2}  = 2   -   2\, \x_i\cdot \x_k  = 2\, \xi_i\cdot\xi_k . \label{56}
\end{equation}

Given three points $\x_1$, $\x_2$ and $\x_3$ of $\sd$ and three
complex numbers $a_1$, $a_2$ and $a_3$, we construct the
rotation invariant quantity
\begin{equation}
J=J(a_1,a_2,a_3) = \left< \left(\Delta \x^2_{12}\right)^{a_3}\left(\Delta \x^2_{23}\right)^{a_1}
\left(\Delta \x^2_{31}\right)^{a_2}\right> \label{invariant}
\end{equation}
where $\left< f\right>$ denotes the average on $\sd$; for three
points
\begin{equation}
\left< f\right> =\frac {\int_{\sd^3}
d\x_1d\x_2 d\x_3 f(\x_1,\x_2,\x_3)}{\int_{\sd^3}
d\x_1,d\x_2, d\x_3 }.
\end{equation}
It follows that
\begin{equation}
J(a_1,a_2,a_3) = {2^{a}\over \omega_{d}^3}
\int (\xi_1\cdot \xi_2)^{a_3}
(\xi_2\cdot \xi_3)^{a_1} (\xi_3\cdot \xi_1)^{a_2} d\mu_{\gamma_0}
(\xi_1)d\mu_{\gamma_0}(\xi_2)d\mu_{\gamma_0}(\xi_3)
={2^{a}\over \omega_{d}^3}
{\hat J}(a_1,a_2,a_3)
\label{defJ}
\end{equation}
and therefore, in view of (\ref{fcJ1}):
\beq
f(a_1,a_2,a_3)
= 2^{-a} \omega_d^3\  c(a_1,a_2,a_3)\ J(a_1,a_2,a_3)
\label{totint}
\endq

In the following sections we will describe the details of the concrete evaluation of  $h_d(\kappa,\nu,\lambda)$.
The method that we present below is based on the previous relations (\ref{fF}) and (\ref{hdw})
applies to all values of the spacetime dimension $d$ ($d\geq 2$).

There exists a remarkable
symmetry relation between of $c$ and $ J$
that follows from symmetry relation
\beq
h_d(\kappa,\nu,\lambda)
=h_d(-\kappa,-\nu,-\lambda)
\label{even}
\endq
which in turn is a consequence  for all real values of $\kappa, \nu, \lambda,$
of the basic symmetry property (\ref{sym}) of the
Legendre functions first-kind.
If we introduce the corresponding
transformation
\beq a_j \to
a'_j = -a_j -{d-1\over 2};
\ j=1,2,3, \endq
 and take Eqs. (\ref{param}) and (\ref{hdw}) into account, we obtain
that for every triplet $(a_1,a_2,a_3)$ such that
$\Re  a_i =\Re  a'_i = -{d-1\over 4}$, the following equality is valid
\beq
f(a_1,a_2,a_3) = f(a'_1,a'_2,a'_3)
\endq
or either (in view of Eq (\ref{totint})):
\beq
2^{-a}{J(a_1,a_2,a_3) \over
c(a'_1,a'_2,a'_3)}
=2^{-a'}{J(a'_1,a'_2,a'_3) \over
c(a_1,a_2,a_3)}
\label{JcJ'c'}
\endq
which are both equivalent to (\ref{even}).

This striking duality
between two integrals over different manifolds is surprising
at first sight.
It will be made clear that the factorization (\ref{totint})
corresponds precisely to a splitting of the
expression (\ref{Theformula}) of
$h_d(\kappa,\nu,\lambda)$ into two parts which are symmetric under that
parity transformation.

In the following section we will evaluate the functions $c$ and $\hat J$.
The study of the integrals (\ref{defF})
and (\ref{hatJ}) will lead us to define and compute the functions
$c$ and $\hat J$ in appropriate domains
of the complex space
${\bC}^3$
of the variables $(a_1,a_2,a_3)$.
$h_d(\kappa,\nu,\lambda)$ is then obtained by taking the restriction of
the holomorphic function $f = c\times \hat J$
to the linear real submanifold $L_d$ of
$\bC^3$
defined by the equations
(\ref{param}), namely
$L_d= \{(a_1,a_2,a_3)\in \bC^3;\ \Re a_j =-{d-1\over 4};\ j=1,2,3\}.$

\section{ Computing $c(a_1,a_2,a_3)$; more on the star-triangle relation}

\label{6}

For computing the integral (\ref{defF}) and obtaining (\ref{st})
with the complete expression of the
function $c(a_1,a_2,a_3) $, we shall consider the following double
Mellin transform:
\begin{eqnarray}
&&\int_0^\infty\int_0^\infty (1+v+z)^{b} v^{s-1}z^{t-1}dv dt=\frac{ \Gamma (s) {\Gamma (t) \Gamma (-b-s-t)}}{\Gamma (-b)};
\end{eqnarray}
this relation is valid for  $\Re(s)>0, \Re(t)>0$, $\Re[b + s+t]
< 0$. Mellin's inversion theorem then provides the following
expansion:
\begin{eqnarray}
(1+v+z)^{b} =-\frac{1}{4\pi^2}\int_{\gamma-i\infty}^{\gamma+i\infty} ds\,v^{-s}
\int_{\gamma'-i\infty}^{\gamma'+i\infty} dt\, z^{-t}  \frac{ \Gamma (s) {\Gamma (t) \Gamma (-b-s-t)}}{\Gamma (-b)}
\end{eqnarray}where the integration paths lie in the strips allowed by the previous inequalities. The formula can be rendered symmetric by homogeneity:
\begin{eqnarray}
(u+v+z)^{b} =-\frac{1}{4\pi^2}\int_{\gamma-i\infty}^{\gamma+i\infty} ds\,
\int_{\gamma'-i\infty}^{\gamma'+i\infty} dt\, u^{b+s+t}v^{-s}z^{-t}  \frac{ \Gamma (s) {\Gamma (t) \Gamma (-b-s-t)}}{\Gamma (-b)}.  \label{22}
\end{eqnarray}

Consider now the expression $(y \cdot \Xi)^{2a}$ where
\begin{equation}
\Xi = \xi_1+ \xi_2+\xi_3,\;\;\;\; a=a_1+ a_2 + a_3;
\end{equation}
all the $\xi_i$'s are lightlike and therefore $\Xi$ is either
timelike or lightlike (in the latter  case $\xi_1,\xi_2$ and
$\xi_3$ lie on the same generatrix of the cone):
\begin{equation}
\Xi^2 = 2\,\xi_1\cdot \xi_2+  2\,\xi_2\cdot\xi_3 + 2\,\xi_3\cdot
\xi_1\geq 0. \end{equation} By application of (\ref{22}) we get that
\begin{equation}
(y\cdot \Xi)^{2a}=  -\frac{1}{4\pi^2}\int_{\gamma-i\infty}^{\gamma+i\infty} ds\,
\int_{\gamma'-i\infty}^{\gamma'+i\infty} dt\,
\frac{{\Gamma (s)  \Gamma (t)  \Gamma (-2a-s-t)}}{\Gamma (-2a)}
{(y\cdot \xi_1)}^{2a+s+t}{(y\cdot \xi_2)}^{-s}{(y\cdot \xi_3)}^{-t}
\label{21a}
\end{equation}
with $\gamma>0,\  \gamma'>0$, $ \ 2\Re(a)+\gamma +\gamma' < 0$.

Let us integrate the two members of (\ref{21a}) over the
Lobatchveski manifold  $\manifold$. The Lorentz invariance of
the l.h.s.  implies that the integral can be parametrized in terms
of the hyperbolic angle $\lambda$ between  $y$  and $\Xi$:
\begin{equation}
\int_{\manifold}(y\cdot \Xi)^{2a}dy = \frac{2\pi^{\frac{d}2}}{\Gamma\left(\frac{d}{2}\right)}(\Xi^2)^{a}
\int_0^\infty(\cosh \alpha)^{2a}(\sinh\alpha)^{d-1} d\alpha = \frac{{\pi^{\frac{d}2}} \Gamma
\left(\frac{1}{2}-a-\frac{d}{2}\right)}{\Gamma\left(\frac{1}{2}-a\right)} (\Xi^2)^{a};
\end{equation}
this result holds provided $\Re (a)  < -(d-1)/2$. Thus
\begin{eqnarray}
&& \frac{2^a{\pi^{\frac{d}2}} \Gamma
\left(\frac{1}{2}-a-\frac{d}{2}\right)}{\Gamma
\left(\frac{1}{2}-a\right)}  (\xi_1\cdot \xi_2+  \xi_2\cdot\xi_3 + \xi_3\cdot \xi_1)^a
 = -\frac{1}{4\pi^2} \frac{2^a{\pi^{\frac{d}2}} \Gamma
\left(\frac{1}{2}-a-\frac{d}{2}\right)}{\Gamma
\left(\frac{1}{2}-a\right) \Gamma (-a)} \times \cr && \times \int_{\delta-i\infty}^{\delta + i\infty} ds\,
\int_{\delta'-i\infty}^{\delta'+i\infty} dt\, (\xi_1\cdot \xi_2)^{a+s+t}(\xi_2\cdot \xi_3)^{-s}(\xi_3\cdot \xi_1)^{-t}
{ \Gamma (t) {\Gamma (s) \Gamma (-a-s-t)}}=\cr &&  = -\frac{1}{4\pi^2}\int_{\gamma-i\infty}^{\gamma+i\infty} ds\,
\int_{\gamma'-i\infty}^{\gamma'+i\infty} dt\,
\frac{{\Gamma (s)  \Gamma (t)  \Gamma (-2a-s-t)}}{\Gamma (-2a)}
\int dy{(y\cdot \xi_1)}^{2a+s+t} {(y\cdot \xi_2)}^{-s}{(y\cdot \xi_3)}^{-t}.
\label{21}\cr &&\label{lastabis}
\end{eqnarray}
In the second step we have applied once more  Eq. (\ref{22})
with $\delta>0,\  \delta'>0$, $ \ \Re(a)+\delta +\delta' < 0$.
Since Eq. (\ref{lastabis}) is valid for any choice of the
vectors $\xi_1$, $\xi_2$ and $\xi_3$ we can multiply $\xi_2$ and
$\xi_3$ by two complex numbers $\alpha$ and $\beta$ and obtain
that
\begin{eqnarray}
&&  \int_{\gamma-i\infty}^{\gamma+i\infty} ds\,\alpha ^{-s}
\int_{\gamma'-i\infty}^{\gamma'+i\infty} dt\,\beta^{-t}
{{\Gamma (s)  \Gamma (t)  \Gamma (-2a-s-t)}}
\int dy{(y\cdot \xi_1)}^{2a+s+t} {(y\cdot \xi_2)}^{-s}{(y\cdot \xi_3)}^{-t}=
\cr
 &&  =  {2^{-a-1}{\pi^{\frac{d-1}2}} } \Gamma
   \left(\frac{1}{2}-a-\frac{d}{2}\right)\times \cr &&
   \int_{\delta-i\infty}^{\delta-i\infty} du\,\alpha^{a+u}
\int_{\delta'-i\infty}^{\delta'+i\infty} dv\, \beta^{-u-v}\,\Gamma (u) {\Gamma (v) \Gamma (-a-v-u)(\xi_1\cdot \xi_2)^{a+v+u}(\xi_2\cdot \xi_3)^{-v}(\xi_3\cdot \xi_1)^{-u}}\cr &&
\label{lastag}
\end{eqnarray}
By changing the variables in the second integral as follows: $u
= -a-s, \; v = a+s+t$. the r.h.s.  becomes
\begin{eqnarray}
\int_{-\Re(a)-\delta-i\infty}^{-\Re(a)-\delta+i\infty} ds\,\alpha^{-s}
\int_{\delta+\delta'-i\infty}^{\delta+\delta'+i\infty} dt\, \beta^{-t}\,
\Gamma (-a-s) {\Gamma (a+s+t) \Gamma (-a-t)}\times \cr \times{(\xi_1\cdot \xi_2)^{a+t}
(\xi_2\cdot \xi_3)^{-a-s-t}(\xi_3\cdot \xi_1)^{a+s}}.\cr &&
\label{lastag2}
\end{eqnarray}
By Mellin's inversion theorem we can now identify the
integrands i.e.
\begin{eqnarray}
&&
\int dy{(y\cdot \xi_1)}^{2a+s+t} {(y\cdot \xi_2)}^{-s}{(y\cdot \xi_3)}^{-t} = \cr &&
\frac{{2^{-a-1}{\pi^{\frac{d-1}2}}} \Gamma
   \left(\frac{1}{2}-a-\frac{d}{2}\right)
 \Gamma (-a-s) {\Gamma (a+s+t) \Gamma (-a-t)}}{\Gamma (s)  \Gamma (t)  \Gamma (-2a-s-t)}(\xi_1\cdot \xi_2)^{a+t}
(\xi_2\cdot \xi_3)^{-a-s-t}(\xi_3\cdot \xi_1)^{a+s} \cr &&
\end{eqnarray}
Finally, by setting $-s = a_3 + a_1$ and $-t= a_1+a_2$  the
proof of the star-triangle relation (\ref{st}) is completed
with
\begin{equation}
c(a_1,a_2,a_3) =  {2^{-a-1}{\pi^{\frac{d-1}2}}
\Gamma\left(-a-\frac{d-1}{2}\right)}
\frac{\Gamma(-a_1)\Gamma(-a_2)\Gamma(-a_3)}{\Gamma(-a_2-a_3)\Gamma(-a_3-a_1)\Gamma(-a_1-a_2)}
\label{cgamma}
\end{equation}

\subsection{Conical limit}
It is interesting to remark that, under certain conditions, a
true star-triangle relation is obtained by integrating over
a cycle of the lightcone. Consider indeed the three-point
function
\begin{equation}
G_{a_1,a_2,a_3}(\xi_1,\xi_2,\xi_3) =
\int_{\gamma_{0}} (\xi\cdot \xi_1)^{a_2+a_3}(\xi\cdot \xi_2)^{a_3+a_1}(\xi\cdot \xi_3)^{a_1+a_2} \alpha(\xi),
\end{equation}
where the integration is performed on the  parabolic basis
$C^+\cap\{\xi^0+\xi^d=1\}$ of the future cone; this corresponds
to setting $\lambda = 1$ in the parametrization
\begin{equation}
 \xi(\lambda,\eta)=\left\{\begin{array}{lcc}
\xi^{0} &=&  \frac\lambda 2 (1+ {\eta}^2)    \\
\xi^{i} &=&  \lambda {\eta^i}\\
\xi^{{d-1}} &=& \frac\lambda 2 (1- {\eta}^2)
\end{array} \right.
\label{abscoor} \;\;\;\;0<\lambda<\infty,\;\;\;\;\eta \in {\mathbb
R}^{d-2},
\end{equation}
and w.r.t. the Lebesgue measure $d\mu(\xi) = d\eta$.
Concretely, since
\begin{eqnarray}  2 \,{\xi(\lambda, \eta) \cdot
\xi'(1,\eta')} = \lambda {\left({\eta}-{\eta'}\right)^2}
 \label{xixicoord}
\end{eqnarray}
\begin{equation}
G_{a_1,a_2,a_3}(\xi_1,\xi_2,\xi_3) =
\frac{\lambda_1^{a_2+ a_3}\lambda_2^{a_3+ a_1}\lambda_3^{a_1+ a_2}}{2^{2a}} \int_{\gamma} [{\left({\eta}-{\eta_1}\right)^2}]^{a_2+a_3}[{\left({\eta}-{\eta_2}\right)^2}]^{a_3+a_1}
[{\left({\eta}-{\eta_3}\right)^2}]^{a_1+a_2} d\eta.
\end{equation}
The result is of course invariant under Euclidean
transformations in $(d-1)$ dimensions. There is however a
special case to be considered: when $2a = (1-d)$  the integrand
is a closed $(d-1)$-form on the future cone. By applying
Stokes' theorem one sees that the integral does not depend on
the choice of the integration manifold and the result is fully
$SO_0(1,d)$-invariant; this invariance can be now interpreted
as Euclidean conformal invariance. As before, by exploiting the
$SO_0(1,d)$-invariance of the integral and the homogeneity
properties of the integrand one has that
\begin{equation}
\int_{\cal M} (\xi\cdot \xi_1)^{a_2+a_3}(\xi\cdot \xi_2)^{a_3+a_1}(\xi\cdot \xi_3)^{a_1+a_2} d\mu(\xi) =  d(a_1,a_2,a_3) (\xi_1\cdot \xi_2)^{a_3}
(\xi_2\cdot \xi_3)^{a_1}(\xi_3\cdot \xi_1)^{a_2}\label{st2}.
\end{equation}
To determine the constant we need to compute the  integral
$\int(\xi\cdot y)^{1-d}\alpha(\xi)$ where again $y =
\xi_1+\xi_2+\xi_3$. This is most easily done using the
spherical basis of the cone $\gamma_0$ with respect to the
measure  $d\mu_{\gamma_0}(\xi)$ (see Eq. (\ref{norms})).
Calculating the previous integral in these coordinates is
immediate:
\begin{equation}
\int_{\gamma_0}(\xi\cdot
y)^{1-d}\alpha(\xi) = \omega_d\, (y\cdot y )^{-\frac{d-1}{2}}
\end{equation}
Now we can follow the same steps as before and obtain the
limiting conformal invariant  star-triangle relation
(\ref{st2}) with
\begin{equation}
d(a_1,a_2,a_3) =  {{(2\pi)^{\frac{d-1}2}}}  \frac{\Gamma(-a_1)\Gamma(-a_2)\Gamma(-a_3)}{\Gamma(-a_2-a_3)\Gamma(-a_3-a_1)\Gamma(-a_1-a_2)}
\end{equation}

\section{Computing $J(a_1,a_2,a_3)$: direct and indirect methods.}
\subsection{Probabilistic interpretation of $J$}
To evaluate $J$ we make use of the standard parametrization of
a point $\x\in\sd$ in terms of $(d-1)$ angles $\theta_i$,
parametrization that we spell here for the reader's
convenience:
\begin{equation}
\x = \left\{ \begin{array}{l}
\cos\theta_1 \\
\sin\theta_1 \cos \theta_2\\
\vdots\\
\sin\theta_1 \ldots \sin \theta_{d-2} \cos \theta_{d-1}\\
\sin\theta_1 \ldots \sin \theta_{d-2} \sin \theta_{d-1}
\end{array} \begin{array}{ll} 0<\theta_1 < \pi, \ldots,  0<\theta_{d-2} < \pi & \\
 0<\theta_{d-1} <2\pi & \end{array}
\right.
\label{sphericalcoord}
\end{equation}
so that
\begin{equation}
d\x = \prod_{j = 1}^{d-1} (\sin\theta_j)^{d-1-j}\ d\theta_j.
\end{equation}
The rotation invariance of (\ref{invariant}) implies that it is
possible to perform  the integrations  by fixing one point, say
$\x_1 = (1,0,\ldots,0)$ and specializing the second point  $\x_2 = (\cos \phi_1, \sin
\phi_1, 0 , \ldots ,0)$ so that
\begin{eqnarray}
\x_1\cdot \x_2  = \cos
\phi_1, \;\;\;\;\; \x_1\cdot \x_3 =
\cos \theta_1,\;\;\;\;\; \x_2\cdot \x_3  = \cos \phi_1 \cos \theta_1 +
\sin\phi_1\sin\theta_1\cos\theta_2.
\end{eqnarray}
It follows that
\begin{eqnarray}
J &=&\frac{4^a\omega_{d-1}\omega_{d-2}}{\omega_{d}^2}\int_0^\pi (\sin\phi_1)^{d-2}d\phi_1
\int_0^\pi(\sin\theta_1)^{d-2}d\theta_1\int_0^\pi(\sin\theta_2)^{d-3}\ d\theta_{2}\times\cr
&& \times \left(\frac{1- \cos\phi_1}{2}\right)^{a_3}
\left(\frac{1- \cos\theta_1}{2}\right)^{a_1} \left(\frac{1- \cos \phi_1 \cos \theta_1 - \sin\phi_1\sin\theta_1\cos\theta_2 }{2}\right)^{a_2}. \label{J}
\end{eqnarray}
Before proceeding with the evaluation of $J$,  let us change
the  integration variables at the r.h.s.  of (\ref{J}) and replace
the angles used there with the distances
\begin{equation}r_1 = \Delta \x_{23},  \;\;r_2 = \Delta \x_{13},  \; \;r_3 = \Delta
\x_{12},\;\;\;\;0\leq r_j\leq 1.
\end{equation}
The Jacobian of the transformation is readily computed:
\begin{eqnarray}
r_1 r_2 r_3\, {d r_{1} \, d r_2 \, d r_3} =  (\sin\phi_{1})^2 \, (\sin \theta_{1})^2 \, \sin \theta_2\, d\phi_1 \, d  \theta_{1} \, d  \theta_{2}
\end{eqnarray}
Since
\begin{equation}
 (\sin \phi_1 \sin\theta_1 \sin\theta_2)^{2} = 2(r_1^2 r_2^2 + r_1^2 r_3^2 + r_2^2 r_3^2)-(r^4_1 + r^4_2 +r^4_3) - r_1^2 r_2^2r^2_3
\end{equation}
it follows that
\begin{equation}
J(a_1,a_2,a_3) =  \int_{D} \rho(r_1,r_2,r_3) \, r_1^{2a_1}\,  r_2^{2a_2} \, r_3^{2a_3}\, dr_1 dr_2 dr_3.
\end{equation}
where
\begin{equation}
\rho(r_1,r_2,r_3) =
\frac{4^a\omega_{d-1}\omega_{d-2}}{\omega_{d}^2}
r_1 r_2r_3[2(r_1^2 r_2^2 + r_1^2 r_3^2 + r_2^2 r_3^2)-(r^4_1 + r^4_2 +r^4_3) - r_1^2 r_2^2r^2_3]^{\frac{d-4}2}_+.
\end{equation}
Below it will be proven that
\begin{equation}
\int \rho(r_1,r_2,r_3)  dr_1 dr_2 dr_3 = 1
\end{equation}
$J(a_1,a_2,a_3)$ are therefore the moments of the probability
density of three random points on $\sd$ constituting a triangle
whose sides have the sizes $r_1$, $r_2$ and $r_3$.

\subsection{Direct evaluation of $J$}
\label{st22}
By introducing a variable $z$ as follows:
\[z= \frac{1- \cos \phi_1 \cos \theta_1}{|\cos \phi_1 -\cos \theta_1|},
\;\;\;\;\;\;\;\sqrt{z^2-1}=\frac{\sin \phi_1 \sin
\theta_1}{|\cos \phi_1 -\cos \theta_1|},\] one can identify
 a well-known integral representation of a Legendre
functions of the first kind  entering at at the r.h.s.  (see e.g.
\cite{bateman1} Eq. 3.7.6):
\begin{eqnarray}
J &=&4^a(2\pi)^{\frac {d-1} 2}
\frac{\omega_{d-1}}{\omega_{d}^2}\int_0^\pi d\cos \phi_1
\int_0^\pi d\cos\theta_1
 \left(\frac{1- \cos\phi_1}{2}\right)^{a_3}\left(\frac{1- \cos\theta_1}{2}\right)^{a_1}\times\cr
   && \times\,
2^{d-3}\left|\frac{\cos \phi_1 -\cos
\theta_1}2\right|^{{d-3}{}+a_2}
\left(\sqrt{z^2-1}\right)^{\frac{d-3}2}
P^{-\frac{d-3}2}_{\frac{d-3}2+a_2}(z).\end{eqnarray} Evaluating
this integral involves several nontrivial steps. In the first
one we change to the variables  $x= (1-\cos \phi_1)/2$ and $y =
(1-\cos \theta_1)/2$ and restrict the domain of integratation
to the region $0<y<x<1$:
\begin{equation}
J_1 = \frac{4^{a+1}{2}^{{\frac{d-3}2-a_2}}(2\pi)^\frac {d-1} 2\omega_{d-1}}{{\Gamma (-a_2)\,\Gamma (\frac{d-1}2+a_2)}\,\omega_{d}^2}
\int_{0}^{1} dx \int_{0}^{x}
dy  x^{a_3}y^{a_1} (x-y)^{{d-3}+a_2}
\int_1^z dw {(z-w)^{-a_2-1}(w^2-1)^{\frac{d-3}2+a_2} };
\end{equation}
in this Equation we have made use of the following remarkable
integral representation of $P^{\mu}_{\nu}(z)$  that will be
established in Appendix \ref{appA}:
\begin{eqnarray}
(z^2-1)^{\frac {d-3}4 }P^{-\frac {d-3}2}_{{\frac{d-3}2+a_2}}(z) = {2}^{-{\frac{d-3}2-a_2}}\int_1^zdw
\frac{(z-w)^{-a_2-1}(w^2-1)^{\frac{d-3}2+a_2} }{\Gamma (-a_2)\;\;\Gamma (\frac{d-1}2+a_2)}.
\end{eqnarray}
In the  second change of variables we replace  $y$ by $\mu x$:
\begin{equation}
J_1  = \frac{4^{a+1}{2}^{{\frac{d-3}2-a_2}}(2\pi)^\frac {d-1} 2\omega_{d-1}}{{\Gamma (-a_2)\,\Gamma (\frac{d-1}2+a_2)}\,\omega_{d}^2}
\int_{0}^{1} d\mu \, \mu^{a_1} (1-\mu)^{{d-3}+a_2} \int_{0}^{1}
dx \,x^{a+d-2} \int_1^z dw \frac{(w^2-1)^{\frac{d-3}2+a_2}}{(z-w)^{a_2+1}};
\end{equation}
$z$ can be expressed in terms of $x$ and $\mu$ as follows
\begin{equation}
z = z_\mu + (1-z_\mu) x, \;\;\; z_\mu = \frac{1+\mu}{1-\mu}.
\end{equation}
Next, we use $z$ as integration variable in place of $x$ (at
constant $\mu$) and we get that
\begin{eqnarray}
&& \int_{0}^{1} dx\,
 x^{a+d-2} \int_1^z dw\,{(z-w)^{-a_2-1}(w^2-1)^{\frac{d-3}2+a_2} } =\cr && =
\left(\frac{1}{z_\mu-1}\right)^{a+d-1} \frac{ \Gamma (a+d-1) \Gamma
   \left(-a_2\right)}{\Gamma \left(a+d-1-a_2\right)}\int_{1}^{z_\mu}dw\,
(z_\mu-w)^{a_1+a_3+d-2}(w^2-1)^{\frac{d-3}2+a_2} .
\end{eqnarray}

By setting  $u=({z_\mu -w})/({z_\mu-1})$ we  obtain the final
expression for $J_1$:
\begin{eqnarray}
 J_1
&=& \frac{4^{a}{2}^{{\frac{3d-5}2}}(2\pi)^\frac {d-1} 2\omega_{d-2}}
{{\Gamma (\frac{d-1}2+a_2)}\,\omega_{d-1}^2}\frac{ \Gamma (a+d-1)}
{\Gamma \left(a_1+a_3+d-1\right)}\times \cr&& \times
\int_{0}^{1}  \,du u^{\frac{d-3}2+a_3} \left( 1-u\right)^{\frac{d-3}2+a_2}
\int_{0}^{u}   v^{\frac{d-3}2 +a_1}
\left(1-v\right)^{\frac{d-3}2+a_2} {dv}
\end{eqnarray}
The integral $J_2$ over the domain $0<x<y<1$ is obtained from
this expression by interchanging $a_1$ and $a_3$. Finally
\begin{eqnarray}
J  &=& J_1+J_2 =\frac{{2}^{{\frac{3d-5}2}+2a}(2\pi)^\frac {d-1} 2\omega_{d-1}}
{{\Gamma (\frac{d-1}2+a_2)}\,\omega_{d}^2}\frac{ \Gamma (a+d-1)}
{\Gamma \left(a_1+a_3+d-1\right)}\times \cr
&& \times\int_{0}^{1}  du\, u^{\frac{d-3}2+a_3}
\left( 1-u\right)^{\frac{d-3}2+a_2}\int_{0}^{1}  {dv}\, v^{\frac{d-3}2 +a_1}
\left(1-v\right)^{\frac{d-3}2+a_2}  =
\cr &=&
\frac{{2}^{{\frac{3d-5}2}+2a}(2\pi)^\frac {d-1} 2\omega_{d-1}}{\,\omega_{d}^2}
\frac{ { \Gamma (a+d-1)}\Gamma \left(\frac{d-1}{2}+a_1\right){\Gamma (\frac{d-1}2+a_2)}\Gamma
   \left(\frac{d-1}{2}+a_3\right)}{\Gamma
   \left(d-1+a_2+a_3\right)\Gamma \left(d-1+a_3+a_1\right) \Gamma
   \left(d-1+a_1+a_2\right)}.
\end{eqnarray}
In the end
\begin{eqnarray}
 h_d &=&  \frac{1}{(2\pi)^{\frac{3d}2}} 2^{-a}\omega_{d}^{2} c(a_1,a_2,a_3) J(a_1,a_2,a_3)\cr &=& \frac{2^{d/2} \Gamma (-a-\frac{d-1}{2})\Gamma (a+d-1)}{(2\pi)^{3/2}
   \Gamma \left(\frac{d-1}{2}\right)}
\prod_{j=1}^3\frac{ { }\Gamma
\left(\frac{d-1}{2}+a_j\right)\Gamma(-a_j)}{\Gamma
   \left(d-1+a_j-a\right)\Gamma(a_j-a)}
\end{eqnarray}
and the formula (\ref{Theformula}) is proven for $d>1$ integer.
Estimates shown in Appendix (\ref{carlson}) show that
interpolation is permitted and the formula is vaild for complex
values of the dimension $d$.
The analyticity in $\kappa$ of the expression at the r.h.s.
also directly confirms our previous statement
the analyticity of $\rho_{\nu,\lambda}(\kappa)$.

\subsection{The $J / c-$duality and an indirect evaluation of  $J(a_1,a_2,a_3)$}
\label{7.1}
There is a possibility to avoid the computation based on the relation (\ref{JcJ'c'}) and on the fact
the integral (\ref{defJ}) defines $J(a_1,a_2,a_3)$ as a holomorphic
function of $(a_1,a_2,a_3)$ in the domain $\{(a_1,a_2,a_3)\in \bC^3;\
\Re (a_1+a_2+a_3) > -(d-1)\}$.

It then follows from the latter property
and from the analyticity properties of the $\Gamma-$functions
involved in the expression
(\ref{cgamma}) of the function $c(a'_1,a'_2,a'_3)$ (with
$a'_j=-a_j-{d-1\over 2}$)
that the l.h.s.  of (\ref{JcJ'c'})
can be analytically continued as a holomorphic function
$l(a_1,a_2,a_3)$
of $(a_1,a_2,a_3)$ in the tube
$T_+ =
\{(a_1,a_2,a_3)\in \bC^3; \Re  a_i > -{d-1\over 3};\ i=1,2,3\}$,\
while the r.h.s. of (\ref{JcJ'c'})
can be analytically continued  as a holomorphic function
$r(a_1,a_2,a_3)$
in the tube
$T_- =
\{(a_1,a_2,a_3)\in \bC^3; \Re  a_i < -{d-1\over 6};\ i=1,2,3\}$.
Note that the two tubes $T_+$ and $T_-$, which are symmetric of
each other with respect to the submanifold $L_d$, have a nonempty intersection
which is a tube containing $L_d$.
Since both sides of (\ref{JcJ'c'}) coincide on $L_d$,
it follows that they both represent the
same analytic function $E(a_1,a_2,a_3)$ in the complex domain
$T_+\cap T_-$, which therefore also admits an
analytic continuation in $T_+\cup T_-$. But in view of the tube theorem,
the holomorphy envelope of
$T_+\cup T_-$ is the full space $\bC^3$, and thereby $E(a_1,a_2,a_3)$ is an
entire function on $\bC^3$.

Moreover, one can show
that the functions $l$ and $r$ are uniformly bounded by a constant $C$
in respective tubes $T'_+$ and $T'_-$ , which are conical open neighborhoods
of the respective "diagonal tubes"
$\delta_+ =
\{(a_1,a_2,a_3)\in \bC^3; \Re  a_i = \alpha > -{d-1\over 4};
\ i=1,2,3\}$ and
$\delta_- =
\{(a_1,a_2,a_3)\in \bC^3; \Re  a_i = \alpha < -{d-1\over 4};\ i=1,2,3\}$.
Since the convex hull of $T'_+\cup T'_-$ is equal to $\bC^3$,
it then follows that
the entire function $E(a_1,a_2,a_3)$ is also uniformly bounded
by $C$ in the whole space $\bC^3$ and is therefore a constant function.
This entails that the l.h.s.  (or r.h.s.) of (\ref{JcJ'c'}) is a constant $\gamma_d$
(only depending on $d$) and that the
integral (\ref{defJ}) is  directly obtained
without additional computation by the formula
\beq
{J(a_1,a_2,a_3)= \gamma_d \ 2^a\
c(a'_1,a'_2,a'_3)}
\ \ ({\rm or}\ \
{J(a'_1,a'_2,a'_3) =\gamma_d\ 2^{a'}\
c(a_1,a_2,a_3)})
\label{Jc'}
\endq
Since $J(0,0,0)= 2^a$,
the constant $\gamma_d$ is easily computed by choosing $a_j =0$ in (\ref{Jc'}),
which yields (in view of (\ref{cgamma})):
\beq
\gamma_d= 2^{3d-5\over 2} \pi^{-{d-1\over 2}}{[\Gamma(d-1)]^2\over
[\Gamma({d-1\over 2})]^3},
\label{gammad}
\endq
and therefore (in view of (\ref{totint})):
\beq
\ h_d(\kappa,\nu,\lambda)=
{\omega_d^3\over (2\pi)^{3d\over 2}\ \omega_{d-1}} \gamma_d
\ c(a'_1,a'_2,a'_3)
\times c(a_1,a_2,a_3)
={ 2^{2d-{7\over 2}}\over \pi^d \Gamma({d\over 2})} c(a'_1,a'_2,a'_3)
\times c(a_1,a_2,a_3)
\endq

\subsection{A corollary}

A formula by Hsu quoted in Szeg\"o's  book  (\cite{szego},  page 390) gives a
weighted integral of a product of three Gegenbauer  polynomials, all having
the same upper index $k= \frac{d-1}2$. That formula is a sort of extension
to  $SO(d)$ of a well known relation established for Legendre polynomials
(upper index $k=1$) and expressing the square of a Clebsh-Gordan coefficient
of the rotation group $SO(3)$.

The K\"all\'en-Lehmann formula given in Eqs. (\ref{KLwick})  and (\ref{mf})
is an identity between two functions  holomorphic in the variables $\zeta$
and $d$. It is possible to perform an analytic continuation of (\ref{KLwick})
to real values of the variable  $\zeta$ belonging to the interval $]-1,1]$
for every $d$.

By studying the behavior at infinity of the integrand one can show that the
integral converges uniformly w.r.t. $\zeta$ in that interval (with the
exclusion of the singular point $\zeta = -1$). Then, it is tempting to
compute the r.h.s.  as an infinite discrete series  of residues (there is
no difference in closing the contour to the right or to the left). But the
so obtained series converges only for $k<1$ i.e. $d < 3$.  The residue of
the pole in $i\kappa = k+n$ (with $k<0$) of $w_\kappa$ is the Gegenbauer
polynomial of degree $n$. By specializing the l.h.s.  of (\ref{KLwick}) to
$i\nu = k+n_1$ and $i\lambda = k+n_2$ the series at the r.h.s.  becomes a
finite sum in the range $|n_1-n_2|<n<|n_1+n_2|$. The aforementioned formula
by Hsu and Szeg\"o is thus a corollary of our result (\ref{Theformula}) via
the  analyticity of the K\"all\'en-Lehmann formula (\ref{KLwick}).

\section{Conclusions and outlook}

The integral (\ref{integral}) that we have studied in this paper gives an
exact evaluation of
the ``phase space'' coefficient in the rate (\ref{1.17}) of the decay process
$1 \to 2$ in a de Sitter universe.
The fact that  $\rho_{\nu,\lambda}(\kappa)$ is an holomorphic function in
their arguments
means that in the de Sitter universe all such decays are possible and there is
no analogue of the mass subadditivity condition of flat spacetime.
This phenomenon should not be ascribed to the thermal features of
the de Sitter universe
\cite{Gibbons:1977mu,Bros:1994dn,Bros:1995js,Bros:1998ik}.

Indeed  the thermal features manifest themselves only when the field algebra
is restricted to a wedge-shaped region bordered by bifurcate Killing horizons
\cite{kayw} while the amplitude (\ref{1.17}) is integrated overall the de
Sitter universe. Second, a similar computation performed in flat space
thermal field theory does not exhibit the same phenomenon.

There several possible other investigation that are rendered possible  by
the results of the present paper.
One is the exploration of the star-triangle relation that we have found and
the associated Yang-Baxter equation.
We will examine that question elsewhere.

\newpage
\appendix
\section{Appendix. Another integral representation for the
associate Legendre functions of the first kind}
\label{appA}

In this appendix we show how to construct the integral representation of
the Legendre function of the first kind that enters crucially in our
computation of the triangular invariant $J(a_1,a_2,a_3)$.

The starting point is the usual expression for
the associated Legendre functions as derivatives
of Legendre polynomials commonly encountered in books of quantum
mechanics:
\begin{equation}
P^{m}_l(x) =  (x^2-1)^{\frac m2}\frac{d^{m}}{dx^{m}}P_l(x) =
\frac{1}{2^l l!}(x^2-1)^{\frac m2}\frac{d^{m+l}}{dx^{m+l}}(x^2-1)^l \,\
\label{legn}
\end{equation}
valid for $x>1$. The idea is to use the Riemann-Liouville fractional
integration operator
\begin{equation}
(D^{\nu} f)(x) = \frac{1}{\Gamma(-\nu)}\int_1^x (x-y)^{-\nu-1}f(y) dy
\end{equation}
to replace the standard  derivative of integer order in (\ref{legn}). The conjectured interpolation formula for the Legendre function of first kind is then given by
\begin{equation}
P^{m}_l(x)  \to \hat P^\mu_\lambda(z) =   \frac{(x^2-1)^{\frac \mu 2}}{2^\lambda\Gamma(\lambda+1)\Gamma(-\mu-\lambda)}
\int_1^z (z-y)^{-\mu-\lambda-1}(y^2-1)^\lambda dy. \label{hypothese}
\end{equation}
To directly evaluate the RHS of (\ref{hypothese}) we set $z-1=2\xi$ with
$-\pi <\Arg \xi < \pi$ and $y-1 = 2\lambda \xi$  with $0\leq \lambda \leq 1$:
\begin{eqnarray}
\hat P^\mu_\lambda(z) &=&    \frac{(2\xi)^{-\mu}(z^2-1)^{\frac \mu 2}}
{\Gamma(\lambda+1)\Gamma(-\mu-\lambda)}
\int_0^1 (1-\lambda)^{-\mu-\lambda-1}\lambda^\lambda (1+ \xi \lambda)^\lambda
d\lambda  \cr
&= & {(z-1)^{-\frac\mu 2}(z+1)^{\frac \mu 2}} \sum_{n=0}^{\infty} \xi^n
\frac {\Gamma(\lambda+n+1)}
{\Gamma(-\mu+n+1)\Gamma(\lambda-n+1)\Gamma(n+1)}\cr
&= & - \frac{(z-1)^{-\frac\mu 2}(x+1)^{\frac \mu 2}}
{\pi}\sin \pi \lambda \sum_{n=0}^{\infty} (-\xi)^n
\frac {\Gamma(\lambda+n+1)\Gamma(-\lambda+n)}
{\Gamma(-\mu+n+1)\Gamma(n+1)} \cr &= &
\frac{(z-1)^{-\frac\mu 2}(z+1)^{\frac \mu 2}}{\Gamma(1-\mu)}
{}_2F_1\left(-\lambda,\lambda+1;1-\mu;\frac{1-z}{2}\right) =
 P^\mu_\lambda(z).
\end{eqnarray}
The last identification follows from \cite{bateman1} Eq. 3.2(14). The proof is
valid in the domain of convergence of the hypergeometric series but the
identification has of course a larger domain of applicability that is the
cut-plane ${\Bbb C} \setminus \{(-\infty,1]\}$ where the integral at the RHS
of (\ref{hypothese}) converges.  A second related  way of evaluating
(\ref{hypothese}) consists in using Mellin's convolutions and  writing
\begin{eqnarray}
 \int_1^x  (x-y)^{-\mu-\lambda-1}(y^2-1)^\lambda dy
= x^{-\mu-\lambda-1}\int_0^\infty
f\left(\frac{x}{y}\right) g(y)\frac{dy}{y}
= x^{-\mu-\lambda-1} F * G(x) \label{conv}
\end{eqnarray}
where
\begin{equation} f(x) =
\left(1-\frac{1}{x}\right)^{-\mu-\lambda-1}_+,\;\;\;g(x)=x(x^2-1)_+^\lambda.
\end{equation}
The Mellin transforms $F(s)$ and $G(s)$ of these functions are readily computed
and their product is the Mellin transform of the convolution (\ref{conv})
\begin{eqnarray}
F(s) G(s)& = & {2}^{\lambda+\mu-1}\, \Gamma (-\mu-\lambda) \Gamma (\lambda+1)\,
\frac{\Gamma \left(-\frac s 2\right)
\Gamma \left(-\frac{s}{2} -\frac{1}{2}-\lambda\right)}
{\Gamma \left(-\frac s2 + \frac{-\mu-\lambda}{2}\right)
\Gamma \left(-\frac s2 + \frac{-\mu-\lambda+1}{2}\right)}
\end{eqnarray}
Inversion of the Mellin transfom can be obtained by applying Slater's theorem
\cite{slater} or either directly by
integrating {\em \`a la} Mellin-Barnes; after a few steps that we do not
reproduce we get again that
\begin{equation}
F*G(x) = \frac{1}{2\pi i} \int^{\gamma+i\infty}_{\gamma-i\infty} F(s)G(s) x^{-s} ds  =  {2^\lambda\Gamma(\lambda+1)\Gamma(-\mu-\lambda)}{(x^2-1)^{-\frac \mu 2}} P^\mu_\lambda(x)
\end{equation}
and the representation is proven to hold true.

There exist of course many other integral representations of Legendre functions.
The advantage of this  specific one  is that it can be used easily for
partial integrations
and is very natural generalization of the classic expression (\ref{legn}).
Strangely enough, we were not able to find this explicitly written in the
form given in
Eq. (\ref{hypothese}) in the literature available to us.

By using the integral representation (\ref{hypothese}) for real $u>1$
we get the following bound:
\begin{equation}
\left | P^{-\frac{d-2}{2}}_{-\frac{1}{2} +
i\nu}(u)\right | \leq \frac{2^{\frac{1}{2}}  \, (u+1)^{ -\frac{\Re d}{4}+ \frac 12} (u-1)^{\frac{\Re d}{4} -1}}{ \left|\Gamma(\frac{d-1}{2} -
i\nu)\, \Gamma(\frac{1}{2} +
i\nu)\right|} \,
\log \left(u+\sqrt{u^2-1}\right)
\end{equation}
so that
 \begin{eqnarray}
\left|\Gamma^3\left(\frac{d-1}2\right) h_d(\n,\nu,\lambda)\right| \le {
{2^{{3\over 2}}  C_d(\kappa,\nu,\lambda) \int_1^\infty
(u+1)^{-{\Re d} + 2} (u-1)^{\frac  {\Re d -5}2}    \left[\log \left(u+\sqrt{u^2-1}\right)\right]^3 du
\over \left |
\Gamma\left({1\over 2}+i\n \right)  \Gamma\left({1\over 2}+i\nu \right)
 \Gamma\left({1\over
2}+i\lambda \right) \right |}
} \cr \label{b.599}
 \end{eqnarray}
 where
 \begin{eqnarray} C_d(\kappa,\nu,\lambda)& = &
\left |{\Gamma^3\left(\frac{d-1}2\right)
\over  \Gamma\left({d-1\over 2}-i\n \right)
 \Gamma\left({d-1\over
2}-i\nu \right)
\Gamma\left({d-1\over 2}-i\lambda \right) } \right | .\label{b.59}
\end{eqnarray}
The change of variable $u = \cosh \phi$ allows to rewrite the integral at the RHS of Eq. (\ref{b.599}) as follows:
\begin{eqnarray}
\int_0^\infty (\phi \coth \phi)^3
\left( \frac{\sinh \frac \phi 2}{\cosh^2 \frac \phi 2}\right)^{ {\Re d -1} }     d\phi \simeq  \frac{2^{-\Re d} }{\sqrt  {\Re d}}
\end{eqnarray}
The asymptotic behaviour at large $d$ of the integral at the RHS has been
estimated
by the the steepest descent approximation; for large $\Re d$ the maximum of
the integrand is located near to $\phi_c = 2 \log \left(1+\sqrt{2}\right)$.
The behaviour of $C_d(\kappa,\nu,\lambda)$ can be obtained from  Stirling's
formula (\cite{bateman1}, p. 47) that tells us that it goes like a constant.
It follows that
\begin{eqnarray}
\left|\Gamma^3\left(\frac{d-1}2\right) h_d(\n,\nu,\lambda)\right|
\le const \frac{2^{-3\Re d/2} }{\sqrt  {\Re d}}
\end{eqnarray}
This asymptotic behavior will be promoted to a bound in the following section.

\section{Appendix. Extension of the main formula to complex $d$}
\label{carlson}
We have shown that the function
$h_d(\n,\nu,\lambda)$ defined in Eq. (\ref{integral})
coincides with
\begin{eqnarray}
 g_d(\n,\nu,\lambda)&=&
\frac{\frac{2^{\frac d 2}}{(4\pi)^{\frac32}
\Gamma\left(\frac{d-1}{2}\right)} \ \prod_{\epsilon=\pm 1}
\Gamma\left(\frac{d-1}{4}
+\frac{i\epsilon \kappa+i\epsilon' \nu+i\epsilon''\lambda}{2}\right)}
{\left [\prod_{\epsilon=\pm 1}\Gamma\left(\frac{d-1}{2}+i\epsilon \n\right)
\right ]
\left [\prod_{\epsilon'=\pm 1}\Gamma\left(\frac{d-1}{2}+i\epsilon' \nu\right)
\right ]
\left [\prod_{\epsilon''=\pm 1}\Gamma\left(\frac{d-1}{2}
+i\epsilon'' \lambda\right)\right ]
}
\cr && \label{b.2}
\end{eqnarray}
for all integer $d \ge 2$ and real $\n$, $\nu$, and $\lambda$.
In this appendix, we will show, by using Carlson's Theorem \cite{titch},
that these two functions coincide wherever they are both defined,
i.e. where the integral in (\ref{legendre}) converges. It is obviously
sufficient to prove this for real values of $\n$, $\nu$, and $\lambda$
satisfying $|\n|< B$, $|\nu|< B$, $|\lambda|< B$ for some arbitrary
$B>0$.

\begin{theorem}[Carlson]
Let $f$ be a function holomorphic in the right half-plane
$\{z\in \bC\ :\ \Re z >0\}$, and satisfying
\beq
|f(z)| \le Ae^{k|z|}\ \ \ \hbox{for all }\ z\ \hbox{with}\ \Re z >0,
\label{b.2.1}\endq
where $A \ge 0$ and $0 \le k < \pi$. If $f(z) = 0$ for
$z=1,\ 2,\ \ldots$, then $f=0$.
\end{theorem}

Recall the formula (\cite{bateman1}, 3.7(1) p. 155)
\beq
P_\alpha^\mu(z) = {2^{-\alpha}(z^2-1)^{-\mu/2} \over
\Gamma(-\mu-\alpha)\Gamma(\alpha+1)}
\int_0^\infty(z+\ch t)^{\mu-\alpha-1}(\sh t)^{2\alpha+1}\,dt,
\label{b.3}\endq
valid for $z \in \bC \setminus (-\infty, 1]$ and
$\Re (-\mu) > \Re \alpha > -1$.
We apply this to the special case
$\Re \alpha = -1/2$, which satisfies the above condition
when $-\Re \mu > -1/2$. We shall suppose $-\Re \mu \ge 0$.
In this case, for $z>1$,
\beqa
|2^\alpha \Gamma(-\mu-\alpha)\Gamma(\alpha+1)||P_\alpha^\mu(z)| &\le&
(z^2 -1)^{-\Re \mu/2} \int_0^\infty (z +\ch t)^{\Re \mu-1/2}\,dt
\nonumber\\
&\le& (z^2 -1)^{-\Re \mu/2} \int_0^\infty (z+1)^{\Re \mu-1/2+\veps}
(\ch t)^{-\veps}\,dt
\nonumber\\
&\le& (z^2 -1)^{-\Re \mu/2}(z+1)^{\Re \mu-1/2+\veps}
\int_0^\infty 2^\veps e^{-\veps t}\,dt
\nonumber\\
&\le& 2^\veps \veps^{-1}\,(z^2 -1)^{-\Re \mu/2}(z+1)^{\Re \mu-1/2+\veps}
\label{b.4}\endqa
This holds for all real $\veps$ such that $0<\veps< -\Re \mu+1/2$,
in particular for $0<\veps<1/2$. Note that setting
$\veps = (2+\log z)^{-1}$ gives a more precise bound, but (\ref{b.4}) will
suffice for our present purpose.
Keeping $\n$, $\nu$, and $\lambda$
real and setting $\mu = -(d-2)/2$, with $\Re d > 1$, this gives
\beqa
|h_d(\n,\nu,\lambda)| &\le&
{2^{{3\over 2}+3\veps}\veps^{-3} \over
\left | \Gamma\left({d-1\over 2}-i\n \right) \Gamma\left({1\over 2}+i\n \right)
\Gamma\left({d-1\over 2}-i\nu \right) \Gamma\left({1\over 2}+i\nu \right)
\Gamma\left({d-1\over 2}-i\lambda \right) \Gamma\left({1\over 2}+i\lambda
\right) \right |} \times\cr
&&\int_1^\infty (z^2-1)^{\Re{d-2\over 2}}\,
(z+1)^{-3\Re{d-1\over 2}+3\veps}\,dz
\label{b.5}\endqa
Bounding $(z^2-1)$ by $(z+1)^2$, the last integral is bounded by
\beq
\int_2^\infty t^{-\Re{d+1\over 2}+3\veps}\,dt =
{2^{-(\Re{d-1\over 2}-3\veps)} \over (\Re{d-1\over 2}-3\veps)}
\label{b.6}\endq
provided $3\veps < \Re(d-1)/2$. We choose e.g. $\veps \le 1/12$.
Thus, for $\Re d \ge 2$,
\beqa
\lefteqn{
|h_d(\n,\nu,\lambda)| \le}\cr
&&{2^{{3\over 2}+6\veps-\Re(d-1)/2}\veps^{-3} \over
(\Re{d-1\over 2}-3\veps)
\left | \Gamma\left({d-1\over 2}-i\n \right) \Gamma\left({1\over 2}+i\n \right)
\Gamma\left({d-1\over 2}-i\nu \right) \Gamma\left({1\over 2}+i\nu \right)
\Gamma\left({d-1\over 2}-i\lambda \right) \Gamma\left({1\over 2}+i\lambda
\right) \right |}.\cr &&
\label{b.7}\endqa

Recall Stirling's formula (\cite{bateman1}, p. 47):
\beq
\Gamma(z) = (2\pi)^\half e^{-z + (z-\half)\log z}\,
\left ( 1+ {z^{-1} \over 12} + {z^{-2}\over 288} + O(z^{-3}) \right ),
\label{st.1}\endq
valid for $z \in \bC \setminus \bR_-$.
In applying (\ref{st.1}), we will ignore the last bracket, at the cost
of introducing an unknown multiplicative constant and supposing
$|z| \ge a > 0$. Thus there is a constant $K$ such that, for $z = x+iy$,
$\Re z >0$ and $|z| \ge a$,
\beq
|\Gamma(z)| \le K|z|^{-1/2} \exp (-x + x\log|z| - {\pi \over 2}|y|),
\ \ \ \
|\Gamma(z)|^{-1} \le K|z|^{1/2} \exp (x - x\log|z| + {\pi \over 2}|y|)\ .
\label{st.2}\endq
If $|\n|<B$, $|\Gamma(\half+i\n)|^{-1}$ is bounded by a constant
(depending on $B$). Assume $\Re d >2$ and $|d-1| > 2e+2B$.
Applying the second inequality in (\ref{st.2}) with
$z=(d-1)/2-i\n$ we find that, for some $K'>0$,
\beq
{1\over \left | \Gamma\left({d-1\over 2}-i\n \right)
\Gamma\left({1\over 2}+i\n \right)\right |} \le
K'|d-1|^{1/2} \exp(\pi|d-1|/4)\ .
\label{b.9.1}\endq
Therefore, for $\Re d >2$, $|d-1| > 2e+2B$, $|\n|<B$, $|\nu|<B$,
and $|\lambda|<B$,
we find that, for some constant $A_1 > 0$ (depending on $B$),
\beq
|h_d(\n,\nu,\lambda)|
\le A_1 |d-1|^{3/2} e^{3\pi|d-1|/4}\ .
\label{b.10}\endq
With the same assumptions on $\n$, $\nu$, and $\lambda$, we
now wish to apply (\ref{st.1}) in formula (\ref{b.2}), taking advantage
of the cancellation of growth which occurs between numerator and
denominator.
Let $z = x+iy = (d-1)/2$. The function $g_d$ has the form
\beq
g_d(\n,\nu,\lambda) = 2^{\frac d 2} (4\pi)^{-\frac32}
\left[ \prod_{j=1}^8 \Gamma\left ({z\over 2}+iu_j\right )\right ]\,
\left[ \prod_{k=0}^6 \Gamma\left (z+iv_k\right )^{-1}\right ]\ ,
\label{b.11}\endq
where $u_j$, $v_k$ are real, $|u_j|<2B$, $|v_k|<B$, ($v_0=0$).
We suppose $x > 1$, $|z| > 4B+e$, $|y| > 4B$. There is a constant $A_2 >0$
such that
\beqa
\lefteqn{
|g_d(\n,\nu,\lambda)| \le A_2 |z|^{-1}}\cr
&&\exp \left [
\left [ \sum_{j=1}^8 {x\over 2}(\log|z| + \log|\half + iu_j|/2|z||-1)
-{\pi\over 2}\left ({|y|\over 2}-2B\right )\right ] +
\right.\cr
&&\left.
\left [ \sum_{k=0}^6 x(-\log|z|-\log|1+iv_k/|z|| +1)
+{\pi\over 2}(|y|+B)\right ]
\right ]\ .
\label{b.12}\endqa
With our assumptions, $\log|\half + iu_j|/2|z|| <0$ and
$-\log|1+iv_k/|z|| <0$, hence
\beq
|g_d(\n,\nu,\lambda)| \le A_2 |z|^{-1}
\exp \left [-3x(\log|z|-1) +{3\pi|y|\over 2} +15\pi B/2\right ],
\label{b.13}\endq
and finally, for some $A_3> 0$,
\beq
|g_d(\n,\nu,\lambda)| \le A_3 |d-1|^{-1} \exp{3\pi|d-1|\over 4}\ .
\label{b.14}\endq
Therefore the function
$d \mapsto h_d(\n,\nu,\lambda) - g_d(\n,\nu,\lambda)$ vanishes by
Carlson's theorem.

\section{Appendix.  Analytic continuation of the K\"all\'en-Lehmann
weight}
\label{rhocont}

In this appendix, we use analytic continuation to obtain the
K\"all\'en-Lehmann weight for a two-point function which
is the product of the two-point functions of two free fields
belonging to the complementary series (a special case was
done in \cite{bem}).
We fix $R=1$, and $d \ge 2$ is an integer.
We have obtained the representation
\beq
w_\nu(\zeta) w_\lambda(\zeta)  =
\int_\bR \kappa \rho_{\nu,\lambda}(\kappa)w_\kappa(\zeta)\,d\kappa\ ,
\label{c.1}\endq
with
\beqa
\lefteqn{
\n\,\rho_{\nu,\lambda}(\n) =
{1\over 2^5 \pi^{d+5\over 2} R^{d-2}
\Gamma\left ( {d-1\over2} \right )}
{\n\,\sh \pi\n \over
\Gamma\left ( {d-1\over2} +i\n \right )\Gamma\left ( {d-1\over2} -i\n \right )}
\times} \cr &&
\times
\prod_{\epsilon,\ \epsilon',\ \epsilon'' = \pm 1}
\Gamma \left({d-1\over 4}
+{i\epsilon\n +i\epsilon'\nu +i\epsilon''\lambda \over 2}
\right )\ .
\label{c.2}\endqa
Eq.~(\ref{c.1}) has been proved for real values of $\nu$ and $\lambda$.
We know however that the l.h.s. of this equation extends to a meromorphic
function of  $\nu$ and $\lambda$ with simple poles at
$i\nu = \pm({d-1\over 2}-n)$ and $i\lambda = \pm({d-1\over 2}-n)$,
where $n \ge 0$ is an integer. As $\nu$ and $\lambda$ become complex,
(\ref{c.1}) will remain valid as long as the r.h.s. can be continued.
The integrand in the r.h.s. is a meromorphic function of $\n$,
$\nu$ and $\lambda$. We denote $\mu = {d-1\over 4}$.
In the variable $\n$, the poles arise from the last product of Gamma
functions.
When their arguments are close to $-n$, where $n$ is any integer $\ge 0$,
these $\Gamma$ functions behave as follows:
\beqa
\Gamma \left( \mu + {i\over 2}(\n +\epsilon'\nu+\epsilon''\lambda) \right )
&\sim&
{2(-1)^n\over n!i(\n-i(2\mu+2n+i\epsilon'\nu+i\epsilon''\lambda))}
\label{u.10}\\
\Gamma \left( \mu + {i\over 2}(-\n +\epsilon'\nu+\epsilon''\lambda) \right )
&\sim&
{-2(-1)^n\over n!i(\n+i(2\mu+2n+i\epsilon'\nu+i\epsilon''\lambda))}
\label{u.11}\endqa
Thus the poles listed in (\ref{u.10}) are the opposites of those in
(\ref{u.11}).

When $\nu$ and $\lambda$ are real, and more generally when
$|\Im \nu|+|\Im \lambda| < 2\mu$,
these poles do not touch the real axis,
so that (\ref{c.1}) holds, by analytic continuation, for all such
values.

For definiteness we first assume $0< \beta= \Im \lambda < \alpha = \Im \nu$.
We also temporarily assume that $\Re \nu \not= 0$, $\Re \lambda \not= 0$,
and $\Re \nu \pm \Re \lambda \not=0$.
The poles are at
\beq
\epsilon\n = -\epsilon'\Re \nu -\epsilon''\Re \lambda
+i(2\mu +2n - \epsilon'\alpha  -\epsilon''\beta),
\label{u.12}\endq
They can be grouped as follows:

\vskip 0.5 cm
\noindent {\bf 1. On the line $-\Re \nu -\Re \lambda +i\bR$} :
\beq
(\epsilon = 1,\ \epsilon' = \epsilon'' =1) \ :\ \ \
\n = -\Re \nu -\Re \lambda +i(2\mu +2n - \alpha  -\beta)\ ;
\label{u.13}\endq
\beq
(\epsilon = -1,\ \epsilon' = \epsilon'' = -1) \ :\ \ \
\n = -\Re \nu -\Re \lambda +i(-2\mu -2n - \alpha  -\beta)\ ;
\label{u.14}\endq
On this line, all poles move down as $\alpha+\beta$ increases,
The mutual distances of these poles remain constant.

\vskip 0.5 cm
\noindent {\bf 2. On the line $\Re \nu +\Re \lambda +i\bR$} :
poles opposite to the preceding.

\vskip 0.5 cm
\noindent {\bf 3. On the line $-\Re \nu +\Re \lambda +i\bR$} :
\beq
(\epsilon = 1,\ \epsilon' = -\epsilon'' =1) \ :\ \ \
\n = -\Re \nu +\Re \lambda +i(2\mu +2n - \alpha  +\beta)\ ;
\label{u.15}\endq
\beq
(\epsilon = -1,\ \epsilon' = -\epsilon'' = -1) \ :\ \ \
\n = -\Re \nu +\Re \lambda +i(-2\mu -2n - \alpha  +\beta)\ ;
\label{u.16}\endq
On this line, all poles move down as $\alpha-\beta$ increases,
The mutual distances of these poles remain constant.

\vskip 0.5 cm
\noindent {\bf 4. On the line $\Re \nu -\Re \lambda +i\bR$} :
poles opposite to the preceding.

\vskip 0.5 cm
When $\alpha+\beta$ increases and reaches $2\mu$, the pole in
(\ref{u.13}) with $n=0$ reaches the real axis, as well as its opposite.
Beyond this, eq. (\ref{c.1}) aquires two discrete contributions
provided by the residues of these poles; these can easily be seen
to be equal. If $\alpha+\beta$ increases further, poles in
(\ref{u.13}) with higher values of $n$ cross the real axis,
contributing new discrete terms to eq. (\ref{c.1}).
In the range $2\mu +2N < \alpha+\beta < 2\mu +2N+2$,
(with integer $N \ge 0$), all the poles in (\ref{u.13}) with $0\le n \le N$
have crossed (as well as their opposites).

Similarly if $\alpha-\beta$ increases so that
$2\mu +2N < \alpha-\beta < 2\mu +2N+2$,
(with integer $N \ge 0$), all the poles in (\ref{u.15}) with $0\le n \le N$
have crossed the real axis (as well as their opposites).

Therefore, if $2\mu +2N < \alpha+\beta < 2\mu +2N+2$ and
$2\mu +2M < \alpha-\beta < 2\mu +2M+2$, where $M$ and $N$ are
non-negative integers, and with our assumptions regarding
$\Re \nu$ and $\Re \lambda$,
\beqa
\lefteqn{
w_\nu(z,\ z')\,w_\lambda(z,\ z') =
\int_\bR \n\,\rho_{\nu,\lambda}(\n)\,w_\n(z,\ z')\,d\n}
\nonumber\\
&&+ \sum_{n=0}^N A_n(\nu,\ \lambda)\,w_{i(2\mu+2n+i\nu+i\lambda)}
+\sum_{n=0}^M B_n(\nu,\ \lambda)\,w_{i(2\mu+2n+i\nu-i\lambda)}\ .
\label{u.17}\endqa
Here
\beqa
\lefteqn{
A_n(\nu,\ \lambda) =
{(-1)^n\over
n!2^2\pi^{d+1\over 2}R^{d-2}\Gamma\left ( {d-1\over2} \right )
\Gamma(-2\mu-2n-i\nu -i\lambda) \Gamma(2\mu+2n+i\nu +i\lambda)} \times}\cr
&&{1\over \Gamma(-2n-i\nu -i\lambda)\Gamma(4\mu+2n+i\nu +i\lambda)}
\ \times\cr &&
\Gamma(-n-i\nu)\,\Gamma(-n-i\lambda)\,\Gamma(-n-i\nu-i\lambda)
\ \times\cr &&
\Gamma(2\mu+n+i\nu+i\lambda)\,\Gamma(2\mu+n+i\lambda)\,
\Gamma(2\mu+n+i\nu)\,\Gamma(2\mu+n)\ ,
\label{u.18}\endqa
and
\beq
B_n(\nu,\ \lambda) = A_n(\nu,\ -\lambda).
\label{u.21}\endq
In the expression for $A_n(\nu,\ \lambda)$,
we note that the poles in $\Gamma(-n-i\nu-i\lambda)$ are cancelled by those
of $\Gamma(-2n -i\nu-i\lambda)$ in the denominator. The poles of
$\Gamma(2\mu+n+i\nu+i\lambda)$ are cancelled by those of
$ \Gamma(2\mu+2n+i\nu +i\lambda)$ in the denominator.
Hence the possible poles of
$A_n(\nu,\ \lambda)$ come from the poles of:
\beq
\begin{array}{llcl}
\Gamma(-n-i\nu)\ :\quad & -n-i\nu = -m &\Leftrightarrow &
\Re \nu =0,\ \ \ \alpha = n-m\ ;\\
\Gamma(-n-i\lambda)\ :\quad & -n-i\lambda = -m &\Leftrightarrow &
\Re \lambda =0,\ \ \ \beta = n-m\ ;\\
\Gamma(2\mu+n+i\lambda)\ :\quad & 2\mu+n+i\lambda = -m &\Leftrightarrow &
\Re \lambda =0,\ \ \ \beta = 2\mu+n+m\ ;\\
\Gamma(2\mu+n+i\nu)\ :\quad & 2\mu+n+i\nu = -m &\Leftrightarrow &
\Re \nu =0,\ \ \ \alpha = 2\mu+n+m\ .
\end{array}
\label{u.19}\endq
Here $m \ge 0$ is an integer.
{}From here on we always assume $0\le \beta < \alpha < 2\mu$, so that
only the first two lines in (\ref{u.19}) remain possible.
This assumption also prevents the occurence of the terms
containing $B_n(\nu,\ \lambda)$ in (\ref{u.17}).
We now consider two special cases.

\subsection{Case $\nu = i\alpha$, $\lambda = i\beta$, $0<\beta<\alpha < 2\mu$}

According to the above discussions, if $N$ is a non-negative integer such
that $ 2\mu +2N < \alpha + \beta < 2\mu +2N+2$,
\beqa
\lefteqn{
w_{i\alpha}(z,\ z')\,w_{i\beta}(z,\ z') =
\int_\bR \n\,\rho_{i\alpha,i\beta}(\n)\,w_\n(z,\ z')\,d\n}
\nonumber\\
&&+ \sum_{n=0}^N A_n(i\alpha,\ i\beta)\,w_{i(2\mu+2n-\alpha-\beta)}\ .
\label{u.22}\endqa
provided neither $\alpha$ nor $\beta$ is an integer. If
$\alpha + \beta < 2\mu$ the formula holds without the $A_n$ terms.
It is easy to check that $\n\,\rho_{i\alpha,i\beta}(\n) \ge 0$.
For $A_n(i\alpha,\ i\beta)$ we find
\beqa
\lefteqn{
A_n(i\alpha,\ i\beta) =
{1\over
n!2^2\pi^{d+1\over 2}R^{d-2}\Gamma\left ( {d-1\over2} \right )
\Gamma(-2\mu-2n+\alpha+\beta)\Gamma(-2n+\alpha+\beta)
\Gamma(4\mu+2n-\alpha-\beta)} \times}\cr
&&
\Gamma(\alpha-n)\Gamma(\beta-n) \Gamma(\alpha+\beta-n)
\Gamma(2\mu-\alpha+n)\Gamma(2\mu-\beta+n)\Gamma(2\mu+n) \times
\hbox to 4cm{\hfill}\cr
&&
{(-1)^n \Gamma(n+2\mu-\alpha-\beta)\over
\Gamma(2n+2\mu-\alpha-\beta)}\ .
\label{u.22.1}\endqa
All factors in this expression except the last fraction are positive
since the arguments of the Gamma functions are positive because of
the conditions $2\mu > \alpha > \beta$ and $\alpha+\beta-2\mu-2n>0$.
The last fraction is of the form
\beq
{(-1)^n \Gamma(n+x)\over \Gamma(2n+x)} = (-1)^n \prod_{q}^{2n-1} (q+x)^{-1}\ .
\label{u.22.2}\endq
The last product contains $n$ negative factors and the result is positive.
The positive sesquilinar form defined on $\SS(X_d)$ by the l.h.s. of
(\ref{u.22}) is thus, in this case, a positive superposition of
positive sesquilinar forms,
each corresponding to an irreducible unitary representation of the
de Sitter group.

It follows that a particle with parameter $i\gamma$ (with $0<\gamma<2\mu$)
can decay into two particles with parameters $i\alpha$ and $i\beta$
(with the preceding conditions satisfied) provided
\beq
\gamma = \alpha +\beta -2\mu-2n,\ \ \hbox{i.e.}\ \
2\mu -\gamma = (2\mu-\alpha) +(2\mu-\beta) +2n,
\label{u.23}\endq
where $n$ is a non-negative integer.

\subsection{Case $\nu = i\alpha$, $0<\alpha<2\mu$,
$\Im \lambda= 0$, $\Re \lambda \not= 0$}

Only a particle in the principal series can decay into two
particles with such parameters. There is no other restriction on
the mass of the decaying particle.

\end{document}